\documentclass[preprint,12pt]{elsarticle}

\usepackage{graphicx}
\usepackage{amssymb}
\usepackage{amsmath}
\usepackage{siunitx}
\usepackage{xspace}
\usepackage{xcolor}
\usepackage{units}
\usepackage{subcaption}
\usepackage{adjustbox}
\usepackage{enumitem}
\usepackage{comment}
\usepackage{adjustbox}

\newcommand{\AEgIS}{$\text{AEgIS}$\xspace}
\newcommand{\antih}{$\:\!\bar{\text{H}}\:\!$\xspace}
\newcommand{\antip}{$\:\!\bar{\text{p}}\:\!$\xspace}
\newcommand{\bias}{\emph{bias}}
\newcommand{\threshold}{\emph{threshold}}

\newcommand{\positroncount}[0]{ a few $10^6$\xspace}

\journal{Nuclear Instruments and Methods in Physics Research A}

\begin{document}

\begin{frontmatter}

\title{A cryogenic tracking detector for antihydrogen detection in the \AEgIS experiment}


\author[vienna]{C.~Amsler}
\author[infn_mi,insubria]{M.~Antonello}
\author[moscow]{A.~Belov}
\author[bs,infn_pv]{G.~Bonomi}
\author[tn,infn_tn]{R.~S.~Brusa}
\author[infn_mi,insubria]{M.~Caccia}
\author[cern]{A.~Camper}
\author[cern]{R.~Caravita}
\author[infn_mi,mi]{F.~Castelli}
\author[lac]{D.~Comparat}
\author[polimi2,infn_mi]{G.~Consolati}
\author[heidelberg]{A.~Demetrio}
\author[ge,infn_ge]{L.~Di~Noto}
\author[cern]{M.~Doser}
\author[cern]{P.~A.~Ekman}
\author[cern,ge,infn_ge]{M.~Fan\`{i}}
\author[polimi1,infn_mi]{R.~Ferragut}
\author[cern]{S.~Gerber}
\author[infn_mi]{M.~Giammarchi}
\author[vienna]{A.~Gligorova}
\author[tn,infn_tn]{F.~Guatieri\corref{corr}}
\ead{fguatier@cern.ch}
\author[vienna]{P.~Hackstock\corref{corr}}
\ead{philip.hackstock@cern.ch}
\author[vienna]{D.~Haider}
\author[cern]{S.~Haider}
\author[cern]{A.~Hinterberger}
\author[mpik]{A.~Kellerbauer}
\author[cern]{O.~Khalidova}
\author[infn_ge]{D.~Krasnick\'y}
\author[ge,infn_ge]{V.~Lagomarsino}
\author[cern]{C.~Malbrunot\corref{corr}}
\ead{chloe.m@cern.ch}
\author[infn_tn,tn]{S.~Mariazzi}
\author[moscow,jinr]{V.~Matveev}
\author[heidelberg]{S.~R.~M\"{u}ller}
\author[infn_pd]{G.~Nebbia}
\author[lyon]{P.~Nedelec}
\author[cern]{L.~Nowak}
\author[heidelberg]{M.~Oberthaler}
\author[cern]{E.~Oswald}
\author[bs,infn_pv]{D.~Pagano}
\author[tn,infn_tn]{L.~Penasa}
\author[prague]{V.~Petracek}
\author[infn_mi]{F.~Prelz}
\author[bo]{M.~Prevedelli}
\author[cern]{B.~Rienaecker}
\author[lac]{J.~Robert}
\author[oslo]{O.~M.~R{\o}hne.}
\author[infn_pv,pv]{A.~Rotondi}
\author[oslo]{H.~Sandaker}
\author[infn_mi,insubria]{R.~Santoro}
\author[bern,fn2]{J.~Storey}
\author[infn_ge]{G.~Testera}
\author[cern]{I.~C.~Tietje}
\author[polimi1,infn_mi]{V.~Toso}
\author[cern]{T.~Wolz}
\author[cern,fn1]{J.~Wuethrich}
\author[mpik,fn3]{P.~Yzombard}
\author[heidelberg2,cern,oslo]{C.~Zimmer}
\author[infn_pv,bs2]{N.~Zurlo}

\address[vienna]{Stefan Meyer Institute for Subatomic Physics, Austrian Academy of Sciences, Boltzmanngasse 3, 1090~Vienna, Austria}
\address[infn_mi]{INFN Milano, via Celoria 16, 20133~Milano, Italy}
\address[insubria]{Department of Science, University of Insubria, Via Valleggio 11, 22100~Como, Italy}
\address[moscow]{Institute for Nuclear Research of the Russian Academy of Science, Moscow~117312, Russia}
\address[bs]{Department of Mechanical and Industrial Engineering, University of Brescia, via Branze 38, 25123~Brescia, Italy}
\address[infn_pv]{INFN Pavia, via Bassi 6, 27100~Pavia, Italy}
\address[tn]{Department of Physics, University of Trento, via Sommarive 14, 38123~Povo, Trento, Italy}
\address[infn_tn]{TIFPA/INFN Trento, via Sommarive 14, 38123~Povo, Trento, Italy}
\address[cern]{Physics Department, CERN, 1211~Geneva~23, Switzerland}
\address[mi]{Department of Physics ``Aldo Pontremoli'', University of Milano, via Celoria 16, 20133~Milano, Italy}
\address[lac]{Laboratoire Aim\'e Cotton, Universit\'e Paris-Sud, ENS Paris Saclay, CNRS, Universit\'e Paris-Saclay, 91405~Orsay Cedex, France}
\address[polimi2]{Department of Aerospace Science and Technology, Politecnico di Milano, via La Masa 34, 20156~Milano, Italy}
\address[heidelberg]{Kirchhoff Institute for Physics, Heidelberg University, Im Neuenheimer Feld 227, 69120~Heidelberg, Germany}
\address[ge]{Department of Physics, University of Genova, via Dodecaneso 33, 16146~Genova, Italy}
\address[infn_ge]{INFN Genova, via Dodecaneso 33, 16146~Genova, Italy}
\address[polimi1]{LNESS, Department of Physics, Politecnico di Milano, via Anzani 42, 22100~Como, Italy}

\address[mpik]{Max Planck Institute for Nuclear Physics, Saupfercheckweg 1, 69117~Heidelberg, Germany}

\address[jinr]{Joint Institute for Nuclear Research, Dubna~141980, Russia}
\address[infn_pd]{INFN Padova, via Marzolo 8, 35131~Padova, Italy}
\address[lyon]{Institute of Nuclear Physics, CNRS/IN2p3, University of Lyon 1, 69622~Villeurbanne, France}
\address[prague]{Czech Technical University, Prague, Brehov· 7, 11519~Prague~1, Czech Republic}
\address[bo]{University of Bologna, Viale Berti Pichat 6/2, 40126~Bologna, Italy}
\address[oslo]{Department of Physics, University of Oslo, Sem SÊlandsvei 24, 0371~Oslo, Norway}
\address[pv]{Department of Physics, University of Pavia, via Bassi 6, 27100~Pavia, Italy}

\address[bern]{Laboratory for High Energy Physics, Albert Einstein Center for Fundamental Physics, University of Bern, 3012~Bern, Switzerland}
\address[heidelberg2]{Department of Physics, Heidelberg University, Im Neuenheimer Feld 226, 69120~Heidelberg, Germany}
\address[bs2]{Department of Civil, Environmental, Architectural Engineering and Mathematics, University of Brescia, via Branze 43, 25123~Brescia, Italy}




\fntext[fn2]{present address: Beams Department, CERN, 1211~Geneva~23, Switzerland}
\fntext[fn1]{present address: ETH Z\"urich, Institute for Particle Physics and Astrophysics, CH-8093 Z\"urich, Switzerland}
\fntext[fn3]{present address: Kastler Brossel Laboratory, Sorbonne Universit\'e, CNRS, ENS Universit\'e PSL, Coll\`ege de France, 4 place Jussieu, case 74, 75252, Paris cedex 05, France}

\cortext[corr]{Corresponding author}

\begin{abstract}

We present the commissioning of the Fast Annihilation Cryogenic Tracker detector (FACT), installed around the antihydrogen production trap  inside the \unit[1]{T} superconducting magnet of the \AEgIS experiment. FACT is designed to detect pions originating from the annihilation of antiprotons. Its 794 scintillating fibers operate at $\SI{4}{K}$ and are read out by silicon photomultipliers (MPPCs) at near room temperature. FACT provides the antiproton/antihydrogen annihilation position information with a few ns timing resolution.\\
We present the hardware and software developments which led to the successful operation of the detector for antihydrogen detection and the results of an antiproton-loss based efficiency assessment. The main background to the antihydrogen signal is that of the  positrons  impinging onto the positronium conversion target and creating a large amount of gamma rays which produce a sizeable signal in the MPPCs shortly before the antihydrogen signal is expected. We detail the characterization of this background signal and its impact on the antihydrogen detection efficiency.
\end{abstract}


\end{frontmatter}


\section{Introduction}
\label{S:Intro}

The Antiproton Decelerator (AD) at CERN hosts several experiments scrutinizing properties of antiprotons or antihydrogen atoms. Precise spectroscopy measurements have already been performed~\cite{ALP172,ALP182} on this most simple form of antimatter atom yielding impressive tests of the CPT symmetry, the combination of the charge, parity and time symmetries. Another focus of research with antihydrogen atoms concerns gravitational interaction of antimatter. To date, only a low precision direct measurement of the effect of the Earth's gravitational field on antimatter has been performed \cite{ALPHAfirstgravity} and several experiments at the AD are planned to perform precise measurements in the coming years~\cite{RSABertsche2018,perez2015gbar, Kell2008}. 
The aim of the \AEgIS experiment is a direct measurement of the force exerted by the Earth's gravitational field on an antihydrogen (\antih) beam~\cite{Deflectometer}. To achieve this measurement \AEgIS needs to produce sub-Kelvin antihydrogen atoms in a pulsed scheme. The chosen mechanism is that of charge exchange~\cite{AegisProposal, PhysRevA.94.022714, Krasnicky_2019} in which a cloud of Rydberg-excited positronium atoms (Ps) are made to collide with a cloud of cold trapped antiprotons (\antip). In this process a Ps loses a positron to the antiproton to produce an \antih atom~\cite{KellerbauerProceeding}. To manipulate and cool \antip plasmas, the \AEgIS apparatus features a complex trap system which culminates in the \antih production Malmberg-Penning trap~\cite{AEgISPbar}, located in the center of a $\SI{1}{T}$ superconductive magnet. After the production trap has been loaded with \antip, a bunch of positrons is shot onto a positron/positronium converter located on top of the trap, the resulting Ps atoms outgoing from the converter traverse the electrodes of the production trap through a grid and interact with the trapped \antip. As of 2018 all of the major components of the \AEgIS experiment have been commissioned and tested~\cite{AEgISPbar, PhysRevA.94.012507, CaravitaProceeding, EXAProceeding} and thus the novel \antih production mechanism could be probed.

The detector dedicated to the detection of \antih at \AEgIS is the Fast Annihilation Cryogenic Tracker (FACT).
The design and required performance of the FACT detector raised several engineering challenges. The requirements of a good vertex resolution and high reconstruction efficiency triggered the construction of a detector enclosed inside the superconducting magnet (to maximize the solid angle coverage and minimize multiple scattering) of the \antih production trap requiring its operation at cryogenic temperatures with a minimal heat dissipation. Achieving a suitable axial position resolution of the annihilating \antih atoms translated in a high detector granularity. A good timing resolution was desirable for diagnostic and detection purposes, including the separation of the signal originating from the annihilation of \positroncount positrons on the target enclosed in the FACT from that of a few \antih atoms annihilating shortly afterwards on the walls of the production trap. 

\section{The \AEgIS antihydrogen detector}
\label{S:detector}

\begin{figure}[htp]
\centering
\includegraphics[width=\textwidth]{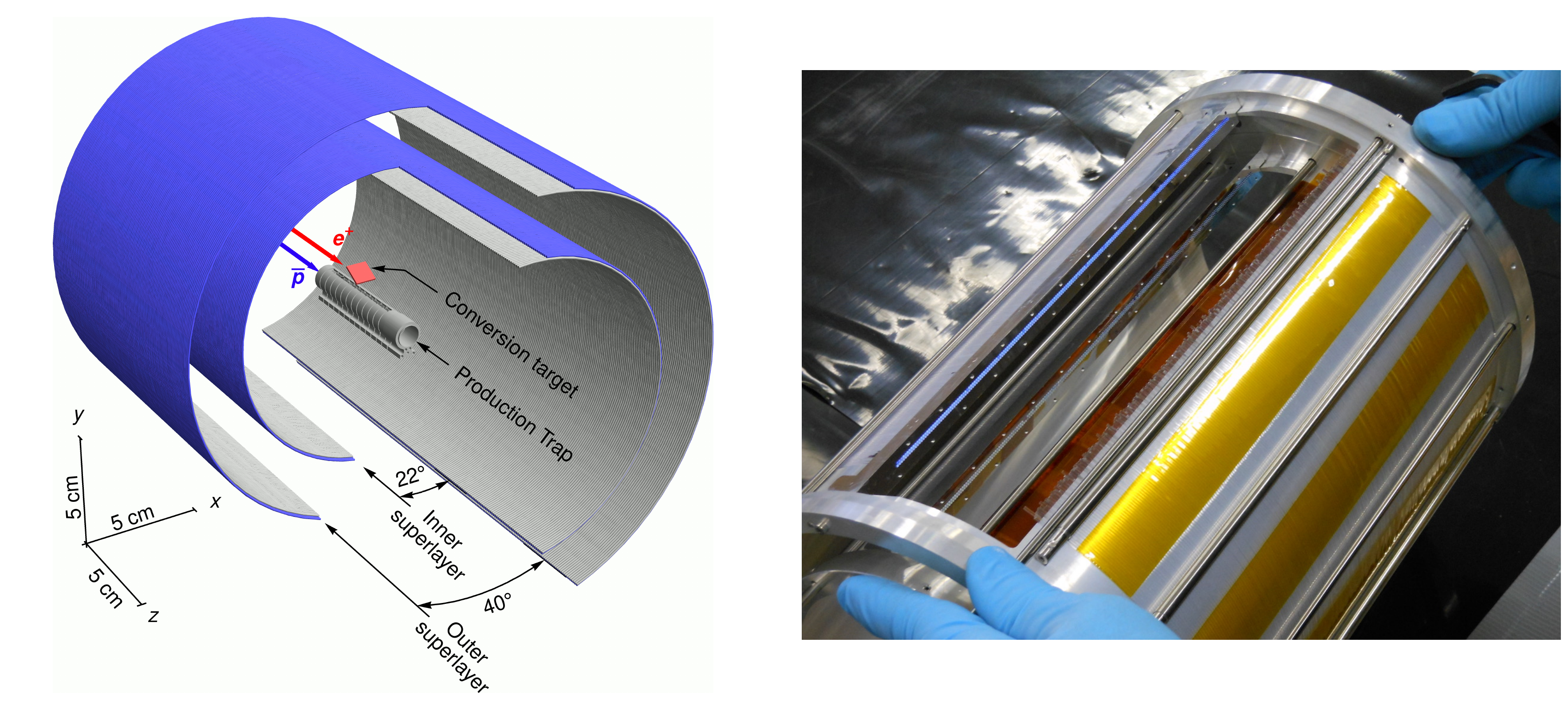}
\caption{Left: cutaway layout of the FACT detector enclosing the \antih production trap. The four layers of scintillating fibers are shown in blue and light gray (the clear fibers connected on one end are not depicted), the antiproton trap is located at the center of FACT, the positron target, depicted in red, is directly above the trap. Right: photograph of the FACT detector before addition of the clear fibers. The detector is rotated with respect to the layout on the left in order to exhibit the gaps thus revealing the two ends of the scintillating fibers.
}
\label{FigureA}
\end{figure}

\begin{figure}[htp]
\centering
\includegraphics[width=\textwidth]{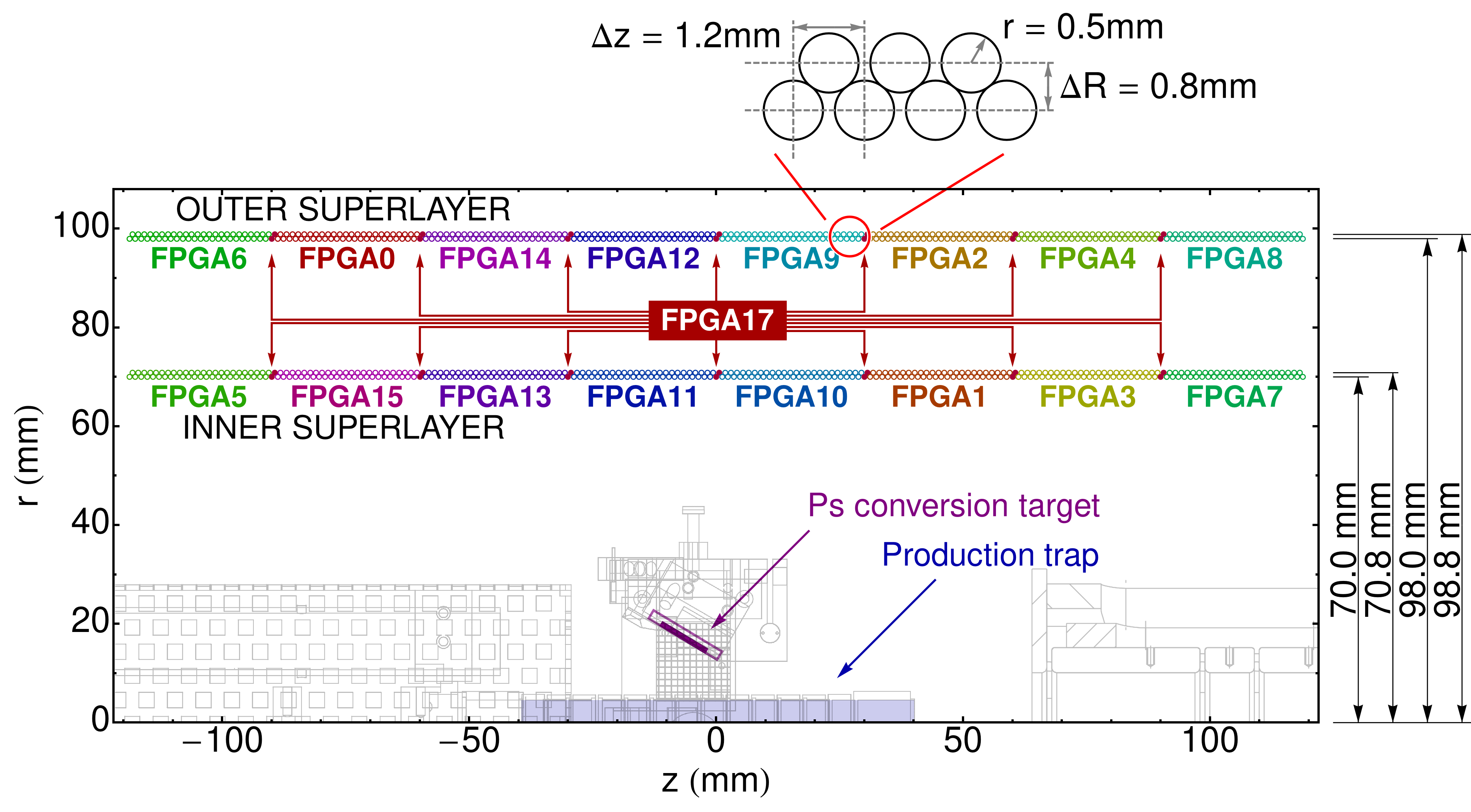}
\caption{Longitudinal projection of the FACT detector showing the two fiber superlayers and the distribution of the detector's fibers among different readout FPGA boards. The origin of the horizontal axis coincides with the center of the FACT detector. To allow for distributed testing of the detector, FPGA 17, which reads a newer generation of MPPCs, was connected to fibers evenly distributed along the direction of the detector's axis.}
\label{FigureB}
\end{figure}

An exhaustive description of the FACT layout has already been provided elsewhere~\cite{AEgISFACT,Storey_2015}, we will recall here its main characteristics.
 Two concentric cylindrical surfaces, which we name \emph{superlayers}, are covered with fibers with respective radii of $\SI{70}{\milli m}$ and $\SI{98}{\milli m}$ as pictured in figure \ref{FigureA}. A total of 794 scintillating fibers (Kuraray SCSF-78 M) with a diameter of \SI{1}{\milli m} ($\unit[6]{\%}$ of which consists of cladding) and a length of either $410$ (inner layer) or $\SI{550}{\milli m}$ (outer layer) are wound around the surface of cylinders that share the production trap's axis and are in thermal contact with a liquid helium cryogenic bath. Each superlayer consists of two layers of fibers wound at radii that differ by $\SI{0.8}{\milli m}$ and staggered along the $z$ axis. Figure~\ref{FigureB} shows a longitudinal view of the two layers of fibers in the superlayers.

\begin{figure}[htp]
\centering
\includegraphics[width=\textwidth]{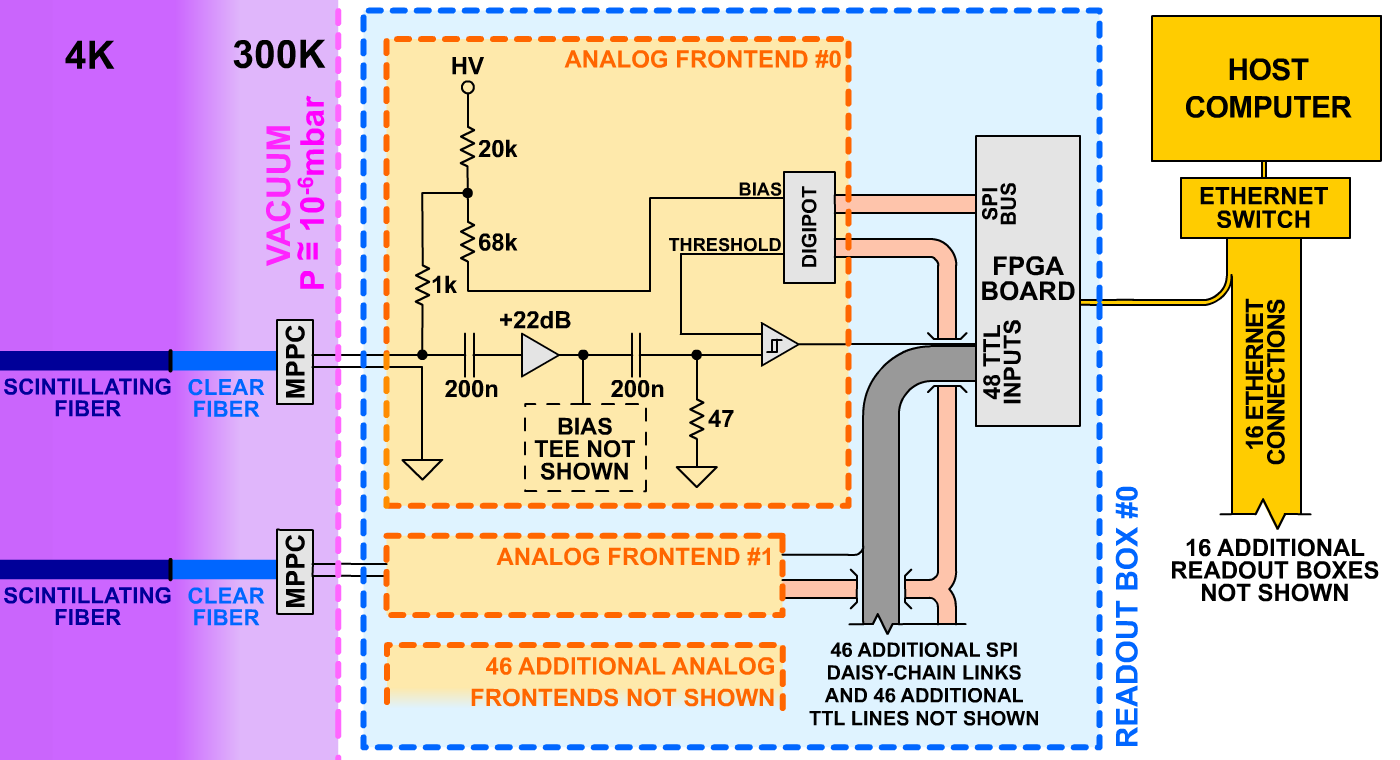}
\caption{General connection scheme of the FACT detector from the fiber readout level to the host computer controlling the detector and performing the readout.}
\label{GeneralScheme}
\end{figure}

Scintillating fibers are coupled with clear fibers which convey photons to light detection sensors. 
In order to accommodate the connection between the scintillating and the clear fibers, the scintillating fibers do not complete an entire revolution around the trap axis: the inner superlayer features a gap of \SI{22}{\degree} and the outer superlayer a gap of \SI{40}{\degree}. Both gaps are located at the bottom of the detector (Fig.~\ref{FigureA}). One end of the fiber is mirror-coated while the other end is connected to a silicon photomultiplier (Hamamatsu S10362-11-100C for most of the detector, 28 fibers employ Hamamatsu S12571-100C photomultipliers instead which exhibit lower afterpulses and a six times smaller dark count rate) called a Multiple Pixel Photon Counter (MPPC). MPPCs are mounted inside the outer vacuum chamber of the AEgIS $\SI{1}{T}$ magnet, on fixtures housing 48 sensors each and featuring a PT-1000 thermal resistor which allows monitoring the MPPC temperatures. The readout electronics is placed outside the vacuum chamber and connected to the thermal sensors and the MPPCs via feedthroughs. The analog signal coming from the MPPCs is amplified and processed by fast discriminators to produce a time-over-threshold signal (ToT). Both the time of the rising and falling edges are recorded by a total 17 FPGA (field-programmable gate array) boards (Xilinx Spartan-6). The FPGA boards are read out through Ethernet connections by a computer. A diagram of the general connection scheme is shown in figure~\ref{GeneralScheme}.

Neutral antihydrogen  formed in the production trap is not  confined and thus eventually reaches the trap wall where it annihilates. During this process the positron annihilates with an electron, generating two \SI{511}{\kilo \eV} $\gamma$ rays, while the antiproton annihilates against a nucleus producing on average $\approx3$ relativistic charged pions behaving as minimum ionizing particles (MIPs) \cite{Hori1}.
The scintillating fibers' detection efficiency for a low energy $\gamma$-ray is of the order of $10^{-3}$ \cite{DavidHaiderMaster};
FACT can thus detect the $\gamma$ burst resulting from the injection of \positroncount positrons in the apparatus but not the two $\gamma$ rays associated with the annihilation of individual \antih atoms. MIPs will however deposit $\sim\unit[200]{keV}$ in a FACT scintillating fiber, leading to around 30 to 50 photons reaching its MPPC \cite{DavidHaiderMaster}.\\

Due to its cylindrical symmetry, its one-sided read-out and given the length of the fibers compared to the time-resolution of the read-out system, the detector is not sensitive to the azimuth position of detected particles. However, making use of the temporal coincidence of events in the two superlayers of the detector, FACT is able to detect the charged MIPs coming from \antip and \antih annihilation and, by reconstructing their point of origin, determine the location of the annihilation vertex along the horizontal axis (see section \ref{S:vertex}).

\subsection{Electronics}

Most of the FACT detector electronics is located inside readout boxes which are attached to the flanges holding the feedthroughs to the MPPC housings. The readout boxes carry out the task of biasing the MPPCs, amplifying and discriminating their signals, recording them using FPGAs and routing the resulting data into specialized Ethernet connections.

The MPPCs are biased from a common $\SI{90}{V}$ source which is reduced to the desired bias voltage through a resistive voltage divider. The resulting bias voltage applied to the MPPCs can be adjusted on a fiber-by-fiber basis between $69.9$ and $\SI{71.1}{V}$ through the first of the two channels of an SPI-controlled digital potentiometer (MCP4242-103E/UN). The resulting voltage source exhibits a high impedance, therefore the voltage on the MPPC terminals is a measure of the current flowing through the sensor. The MPPC voltage is capacitively coupled with a monolithic amplifier (MAR-6+), then digitized through a fast discriminator (ADCMP601) that compares it to a reference voltage, generated through the second channel of the digital potentiometer.
A MIP will produce around 2000 photons in a fiber. Taking into account the trapping efficiency of the fiber (\unit[5.4]{\%}), the loss at the interface between the clear fiber and the MPPC (around \unit[50]{\%})and the MPPC detection efficiency (around \unit[35]{\%}), this will result in a typical \unit[15-25]{photoelectrons} signal at the MPPC. After amplification this signal corresponds to $\sim\unit[120]{mV}$ depending on the particular MPPC gain settings \cite{DavidHaiderMaster}.
The discriminated signal is routed to the board holding the readout FPGA chip. Every board is capable of reading out 48 digital channels with a temporal resolution of $\SI{5}{\nano s}$. To synchronize the acquisition across all of the boards, a TTL trigger signal ($S_{trig}$), generated by the \AEgIS trap control system, is fanned out and distributed to all of the boards. To ensure that synchronization of the FPGAs is maintained during the course of long acquisitions (which can last several seconds) a master clock signal is generated and daisy-chained across all boards.

\subsection{Interface}
The interface follows the Ethernet standard. In order to ease the synthesis of the FPGA code, we forewent the encapsulation of the traffic even to the transport layer 
(we adopted no packet encapsulation)
and opted, instead, to have the control computer and the FPGAs exchange raw ethernet frames.
The data recorded by FACT's FPGAs comes in 12-byte long words including a checksum byte.
To reduce the volume of data that needs to be transferred, we saved the fiber status only when at least one of the 48 inputs changed.
To facilitate the FPGA synthesis we opted to design the control packets so that their fields are 12-byte aligned like the data words are and preserved the same checksum byte in the control packet structure too.

\subsection{Control software}

Control and readout of the FACT detector are performed by a host computer running Linux. The control and readout software, named \emph{FACTDriver}, was developed in C++ employing solely POSIX APIs~\cite{POSIX} and the standard libraries. During the course of each execution of \emph{FACTDriver}, the program performs three tasks:

\begin{enumerate}
	\item It initializes the detector, operation which includes a quick connection test of each FPGA, setting the \bias{} and \threshold{} value for each fiber in the detector and arming the detector. After the detector is armed, \emph{FACTDriver} waits for $S_{trig}$.
	\item When the trigger is received, \emph{FACTDriver} downloads the raw data from the FPGAs through the Ethernet connection into the host computer's RAM. This operation can be performed either asynchronously,	
	with the data transfer happening after the acquisition has concluded or synchronously, with the data transfer being initiated as soon as the trigger is received and the data being streamed while the acquisition is still ongoing.
	\item When the data transfer is concluded, \emph{FACTDriver} decodes the raw data coming from the detector into the format employed by \AEgIS to store it, then transfers it to the experiment's centralized data acquisition system~\cite{Prelz}.
\end{enumerate}

The implementation of \emph{FACTDriver} was heavily optimized to handle the network interface efficiently, with particular attention on the prompt emptying of the input buffer of the Ethernet interface, resulting in a typical input packet handling time below $\unit[1]{\mu s}$. This is necessary in order to minimize the chance that packets could be dropped, and thus the quality of service (QoS) be degraded, when the detector readout is performed synchronously.  We achieved to transfer data without losses from the detector to the host computer with a rate of $50000$ state changes per fiber per second which is comfortably beyond what is required by \AEgIS.

It is worth mentioning a small adjustment that greatly increased the QoS of \emph{FACTDriver}: in order to best take advantage of the eight cores of the host machine, \emph{FACTDriver} was structured as a multithread program, with a thread dedicated to each FPGA board and two to three auxiliary threads. We observed that by staggering the launch of the threads by $\SI{20}{\micro s}$ we reduced drastically the packet collision rate, bringing it from about 20 per \emph{FACTDriver} execution cycle to less than 1.

\section{Equalization}
\label{S:Equalization}

\begin{figure}[htp]
\centering
\includegraphics[width=\textwidth]{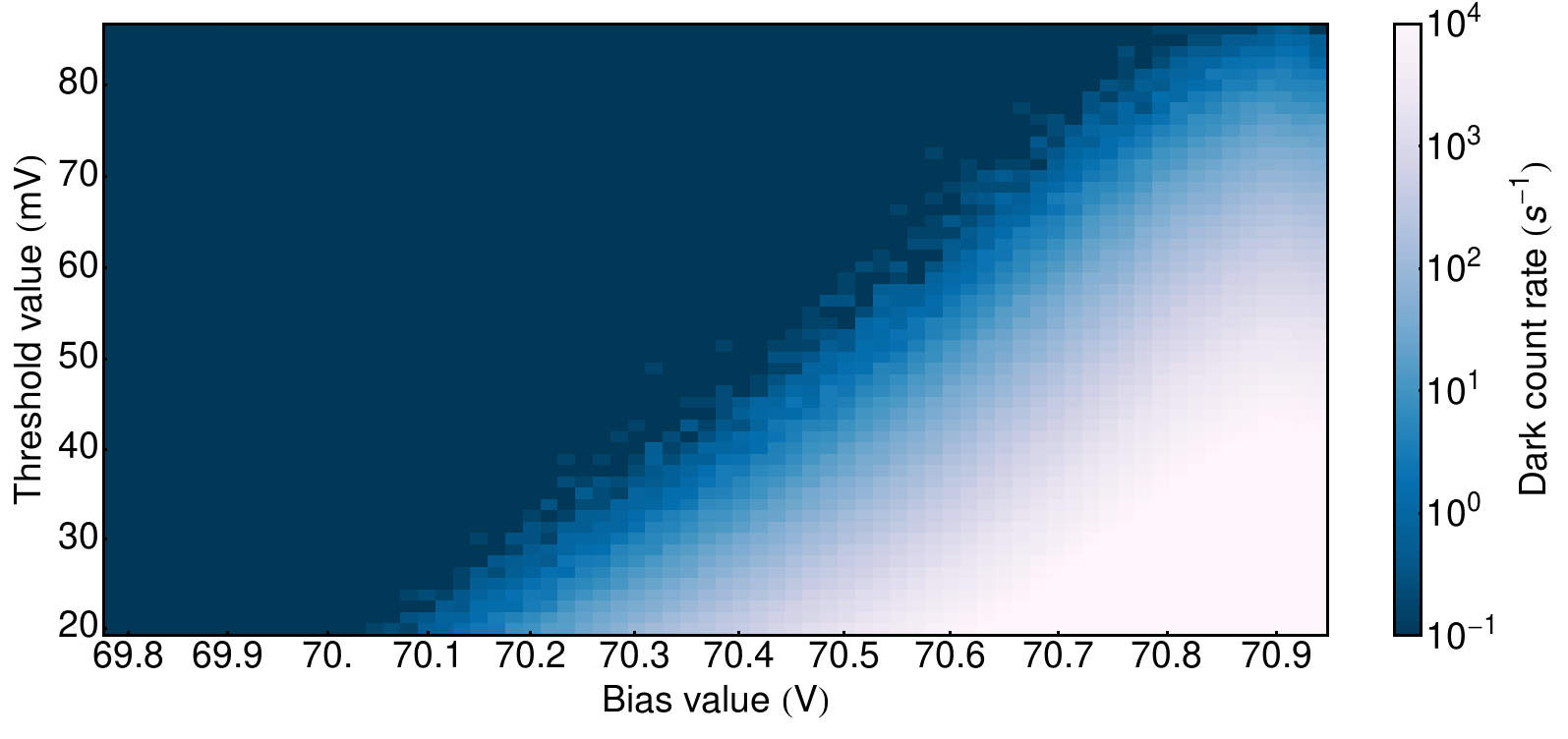}
\caption{Dark count rate of the 24th fiber of FPGA 8 measured across the entire spectrum of bias and threshold values available. It can be seen that although the dark counts can be regulated by using either parameter, the \bias{} has the largest range.}
\label{FigureC}
\end{figure}

The sensitivity and efficiency of the scintillating fibers in detecting pions are determined by two parameters that can be tuned on a fiber-by-fiber basis: the MPPC biasing voltage and the discrimination threshold employed in the analog-to-digital conversion. Due to the low electronic noise present in the acquisition chain, the effects of the two parameters are similar. The former determines the analog signal amplitude, the latter the threshold at which it is discriminated. A change in any of those two parameters will affect the needed amount of energy deposited in the fiber to trigger an output digital signal. Figure \ref{FigureC} shows the dependence of the dark count rate on the settings of both these parameters.
Since between the \threshold{} and \bias{} controls the latter holds the largest effect range, we opted to fix the \threshold{} parameter and then to determine the \bias{} values that are necessary to produce a given dark count which we establish a priori. The goal of the equalization procedure is to have the same dark count rate throughout the detector.

\begin{figure}[htp]
\centering
\includegraphics[width=\textwidth]{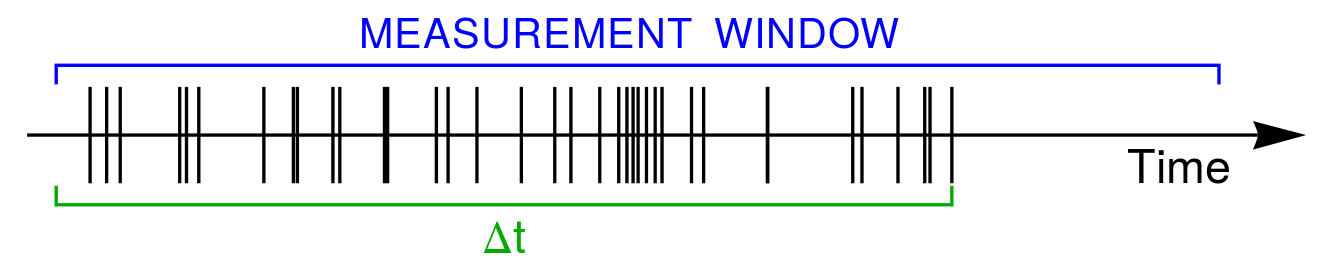}
\caption{A saturating buffer in an FPGA will show, at the end of the recording, a buffer whose end is devoid of events. We therefore choose the observation time $\Delta t$ to be the portion of the buffer ranging up until the last observed event in order to prevent incorrect assessments due to saturation.}
\label{FACTEqualizationScheme}
\end{figure}

Our first attempt to perform the detector equalization was based on an iterative bisection procedure on  the \bias{}. This method is very effective, and can be used to determine the 794 \bias{} values in a matter of minutes. The main shortcoming of such an approach is that a significant amount of observation time is used in sampling points lying far off from the expected setting. Moreover, the algorithm is not robust against saturation of the FPGA buffer during the data acquisition and saturation is very likely to occur at the moment in which high bias values are being scanned/observed. Overcoming this shortcoming is a complex issue, since the consequences of the saturation of an FPGA buffer affect all 48 fibers acquired by that FPGA; therefore, to safely run the ``bisection'' algorithm only one fiber per FPGA can be calibrated at a given time, resulting in a long equalization time. To overcome these difficulties we developed a more efficient algorithm to perform the FPGA equalization which is detailed below.\\
First we describe a procedure to determine the dark rate of a FACT fiber when subjected to a certain \bias{} setting. To perform a rate observation a given \emph{Bias} level is set,  FACT records noise for a certain interval of time at the end of which the number of rising edges recorded in the given time interval is counted. It is however possible for the FPGA event buffer to saturate during the observation time, since for the equalization procedures we do not employ the data streaming capability. In a buffer saturation scenario all of the events following the buffer saturation are dropped. Let $\Delta t$ be the time between the start of the recording and the time of arrival of the last recorded event and $n$ be the number of recorded dark events (Fig.~\ref{FACTEqualizationScheme}). If $n \neq 0$ we can write an estimate of the fiber's dark rate $R$ as

\begin{eqnarray}
	R ~=~ \frac{n \pm \sqrt{n}}{\Delta t}, 
\end{eqnarray}

while if no event was recorded we will estimate it as:

\begin{eqnarray}
	R ~\leq~ \frac{1}{\Delta t},
\end{eqnarray}
$\Delta t$ being in this case the full acquisition time window.\\
It is possible to combine together multiple observations by adding the observation times and the event counts; which results in general in an improvement of the precision of the rate assessment.

Employing an arbitrary start time does not affect the measurement, as the interval between the arbitrary start and the first event obeys the same distribution law as all other time intervals between successive events.
Instead, the time elapsing from the last recorded event to the end of the observation window is not independent of the recorded data (consider e.g. the case of buffer saturation); this is the reason why we defined as $\Delta t$ only the interval ranging up until the last recorded event.

As long as this procedure is strictly applied, we can measure at once the dark count rate of all fibers in the detector. This is allowed as the effect of a noisy fiber filling up its corresponding FPGA buffer will at most be that of reducing the observation time for all the fibers connected to the same FPGA, but will not induce any incorrect assessment of rates.

\begin{figure}[htp]
\centering
\includegraphics[width=0.8\textwidth]{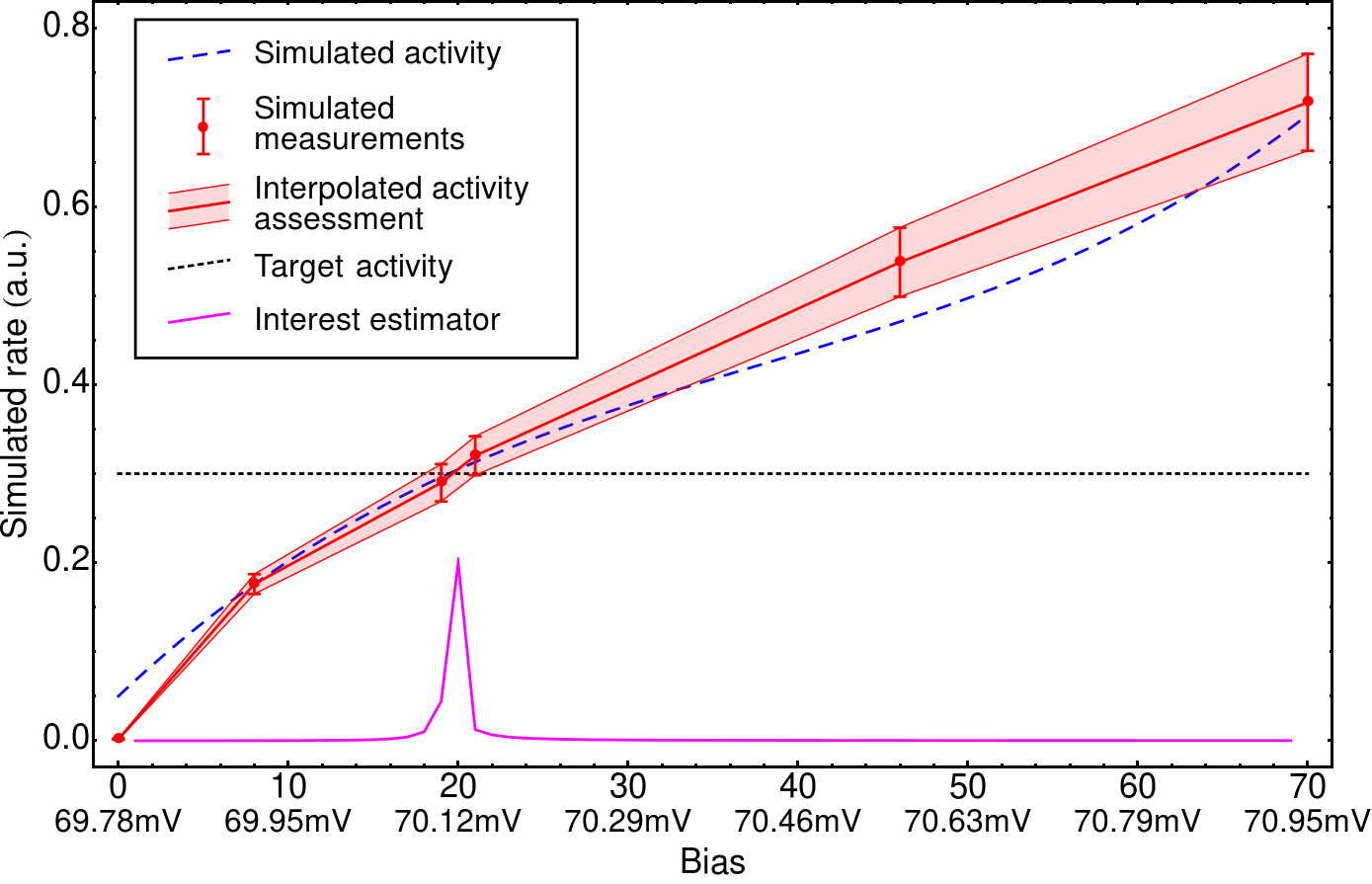}
\caption{An example of a simulated equalization procedure for a single fiber (here we display its sixth iteration). The blue dashed line shows the simulated fiber activity which in a real case scenario we observed to be much steeper. The red envelope shows the interpolated assessment of $R$ and the horizontal black-dotted line indicates the targeted dark noise activity $T$.  In magenta, the interest estimator $I$ (employing an arbitrary scale) in shown indicating that it is already strongly peaking at the optimal \emph{Bias} setting.}
\label{SmartEqualization}
\end{figure}

To complete the description of an equalization procedure for the entire detector we need now to determine how \emph{Bias} values are chosen before each iteration of the algorithm.  
During the first two iterations we measure the rate given by the lowest and the highest available \emph{Bias} values. From the third iteration onward the process becomes more complex. We first give an estimate of the fiber dark count rate for each possible value of the \emph{Bias} by linearly interpolating the mean and uncertainty of the measurements.
We then associate to each possible integer \emph{Bias} setting $b$ an estimator $I(b)$ of how useful it would be for us to measure the dark count given by $b$. 
Originally we intended $I(b)$ to be computed as the probability, given the current measurements, that $b$ would be the optimal \emph{Bias} setting.
Instead, by testing the algorithm in simulated scenarios, we found that a much better performing $I(b)$ is given by the following quadratic law:
\begin{eqnarray}
	I(b) ~=~ \left(\, \frac{\Delta R(b)}{R(b) \,-\, T} \,\right)^2,
\end{eqnarray}
in which $R(b)$ is the rate assessment for the \emph{Bias} value $b$, $\Delta R(b)$ is the uncertainty on $R(b)$, and $T$ is the goal dark noise activity for the specific fiber. 
The simulations carried out to test the algorithm consisted in selecting a monotone response to the \bias{} setting from a wide family of functions, then running the equalization algorithm multiple times generating every time with a Poisson distribution the number of rising edges recorded by the fiber in the given observation time.\\
At each iteration of the algorithm and for each fiber we determine which \emph{Bias} setting $b_{max}$ maximizes $I(b)$; then randomly pick the \bias{} value to be measured among $b_{max}$, $b_{max}+1$ and $b_{max}-1$. This random choice is necessary since, as our simulations showed, it is possible for the best \emph{Bias} setting to temporarily appear farther away from the target activity than another sub-optimal \emph{Bias} value. When this happens the optimal value might not be measured ever again and thus the algorithm may converge onto a sub-optimal value. The aforementioned randomness prevented this from happening in every simulated scenario.
Figure \ref{SmartEqualization} shows an ongoing simulated equalization process in which both the interest estimator $I(b)$ and the interpolated dark rate assessment $R(b) \pm \Delta R(b)$ are displayed.

This algorithm has shown excellent performances both in simulated contexts and when applied to the detector. On FACT the algorithm typically requires less than 12 iterations to converge to the best setting; as a precaution we decided to have the algorithm run 24 iterations during a typical detector equalization. With this procedure, the entire equalization process requires less than $\SI{4.6}{s}$ to be executed, which allows for frequent re-equalization of the detector. 

The equalization procedure can fix the desired dark count rate on a per-fiber basis; for most operations we found setting it at $\SI{50}{s^{-1}}$ a good compromise between noise and efficiency. The possibility to equalize the dark count rates before every run of \AEgIS allows us to increase the reproducibility of measurements and to safely work with higher dark count rates. Frequent re-equalization ensure that any degradation of the gain of the MPPCs caused by thermal drifts  will not have time to take place before the next equalization procedure is performed.

\section{Temperature control}

Many operational parameters of MPPCs, such as dark counts and efficiency, depend on the operation temperature and are well documented~\cite{DinuMPPC}; nonetheless estimating the impact of this dependency on the operation of FACT has been a non trivial task. The introduction of an equalization procedure which can be run in a time scale considerably shorter than that of the thermal excursions to which the detector is subjected (mainly as a consequence of the day/night cycle and of cryogenic servicing of the FACT \SI{1}{T} magnet) has prompted the interest to investigate the dependency of the \bias{} values on the temperature.

\begin{figure}[htp]
\centering
\includegraphics[width=\textwidth]{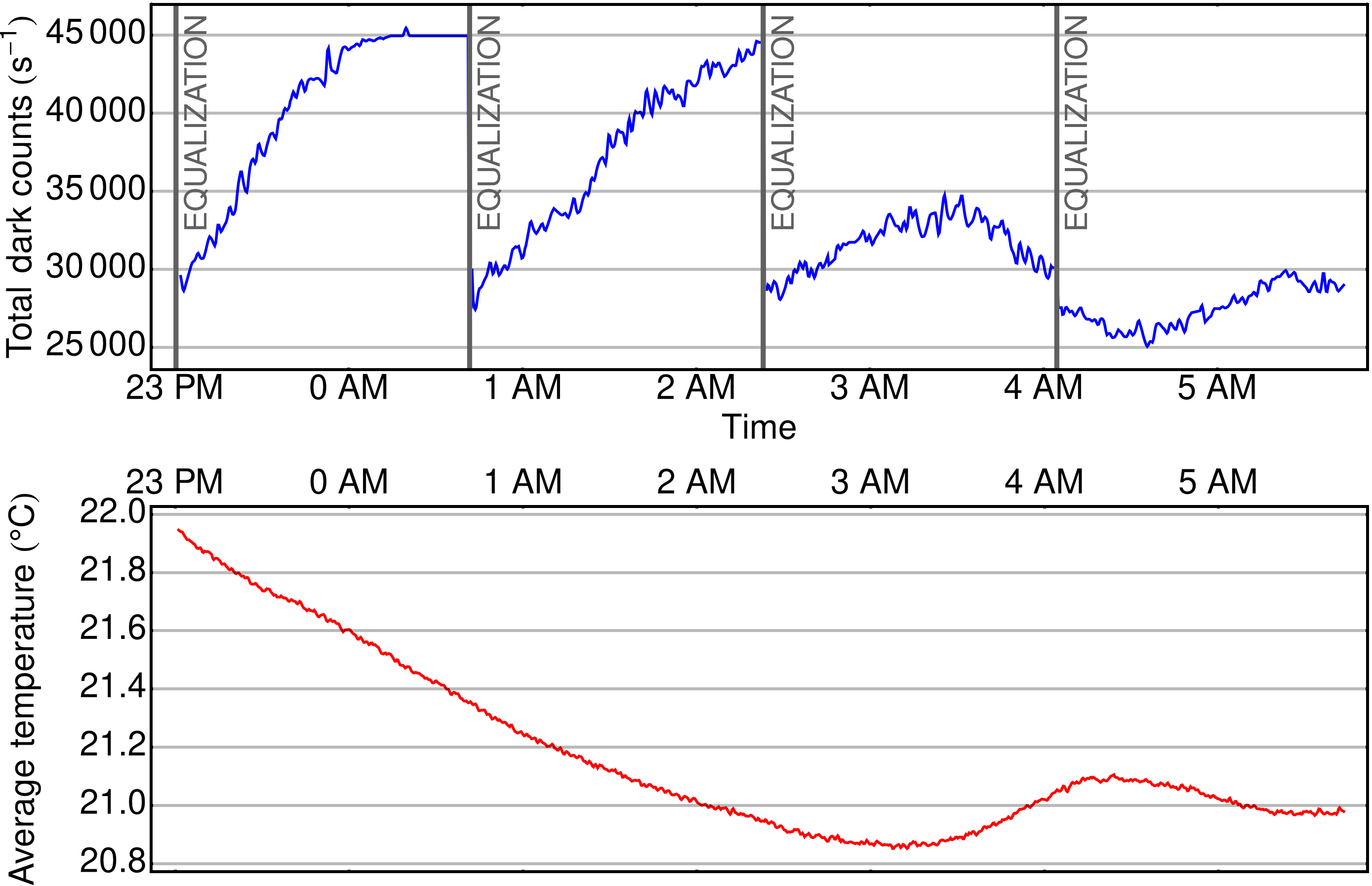}
\caption{Above : the total dark count rate of the FACT detector, monitored over 6 hours. Every 100 minutes the detector equalization was repeated, bringing, for this particular test, the single-fiber dark count rate to \SI{35}{s^{-1}}. Below : the average detector temperature monitored over the same time span; the drift in the dark count rate induced by the temperature variation is evident.}
\label{DarkCountT}
\end{figure}

\begin{figure}[htp]
\centering
\includegraphics[width=0.75\textwidth]{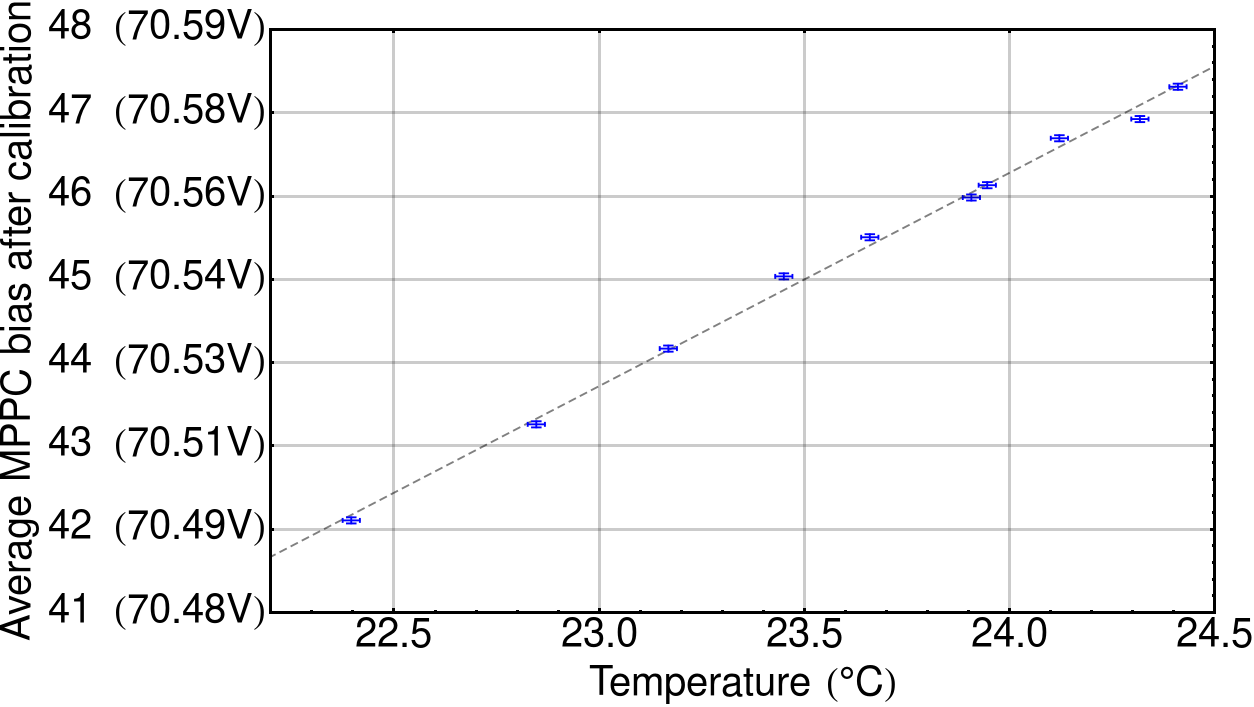}
\caption{Dependency of the mean \emph{Bias} determined by the equalization procedure, as a function of the average MPPC temperature. The measurement has been performed by periodically calibrating the detector in the span of a night.}
\label{BiasT}
\end{figure}

Figure \ref{DarkCountT} shows the total dark count rate of FACT monitored during the span of an entire night. Every 100 minutes the equalization procedure was run to avoid the thermal drift to bring the detector to saturation. As pointed out in section \ref{S:detector}, each read-out box is equipped with a temperature sensor: the drift of the MPPC average temperature during the night is plotted along that of the dark count. The effect of the temperature on the dark count rate is clearly seen by comparing the two plots. We can fix the dark count rate by periodically equalizing the detector, in which case a change in temperature will induce a variation of the \emph{Bias} levels. Figure \ref{BiasT} shows the dependence of the average bias level determined by the equalization procedure as a function of the average detector temperature; this dependency can be approximated as linear in the investigated range.

The MPPCs of FACT are subject to two thermal baths: the cryogenic bath of the superconducting magnet (towards which they are to a certain extent insulated) and the analog frontend board with which they have a better thermal contact. The FACT frontend operates typically at a temperature higher than ambient due to the heat generated by the frontend electronics. To prevent overheating, each frontend box is equipped with a cooling fan, each of which is individually driven to provide a temperature control.

We implemented the MPPC temperature stabilization as a closed-loop proportional--integral--derivative (PID) control. Ideally PID would require the continuous control of the fan rotation speed; this is not feasible with our hardware as our fans cannot turn slower than a fixed minimum. Instead, we took advantage of the fact that the timescale at which the detector temperature drifts is long enough to employ a slow pulse--width--modulation (PWM) control with a pulse period of one minute to reduce the cooling power. The readout as well as the fan control being available on a per-box basis allowed us to control the temperature of the MPPC fixtures individually.

\begin{figure}[htp]
\centering
\includegraphics[width=\textwidth]{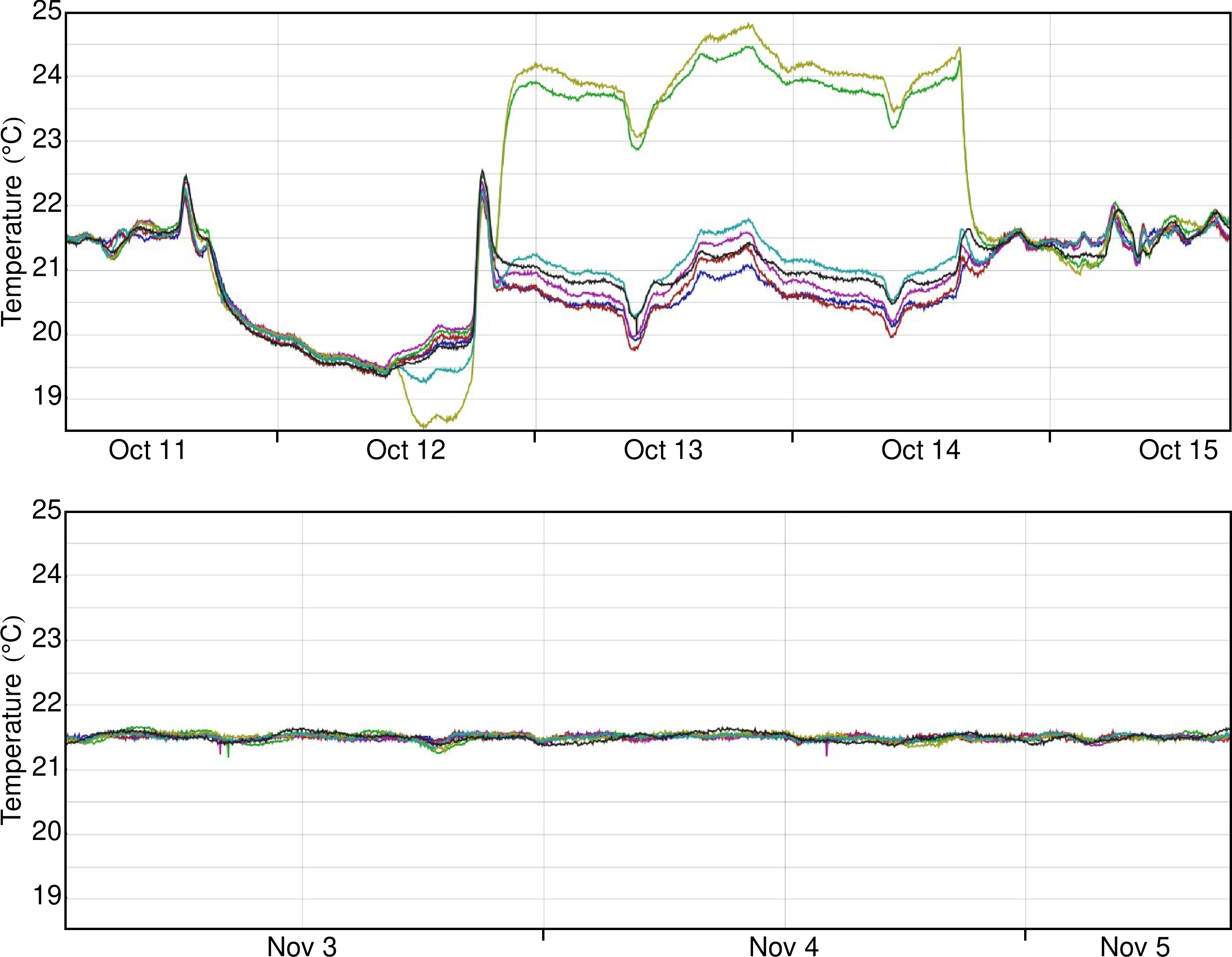}
\caption{Upper: temperature excursion of the MPPC boards with the thermal control disabled. Lower: the temperature excursion observed with the thermal control enabled.}
\label{TControl}
\end{figure}

Figure \ref{TControl} shows the thermal drift of the MPPC holders during the span of several days both with and without active temperature control. Under active temperature control the excursion is reduced to $0.06\,^\circ \text{C}$ RMS, $0.3\,^\circ \text{C}$ peak-peak (0.77 in \bias{} setting units).
The combination of the temperature control and frequent equalization of the detector ensures the maximum stability of the measurement conditions which is important for the quality of the combined data set.

\section{Vertex reconstruction}
\label{S:vertex}

\begin{figure}[htp]
\centering
\includegraphics[width=0.8\textwidth]{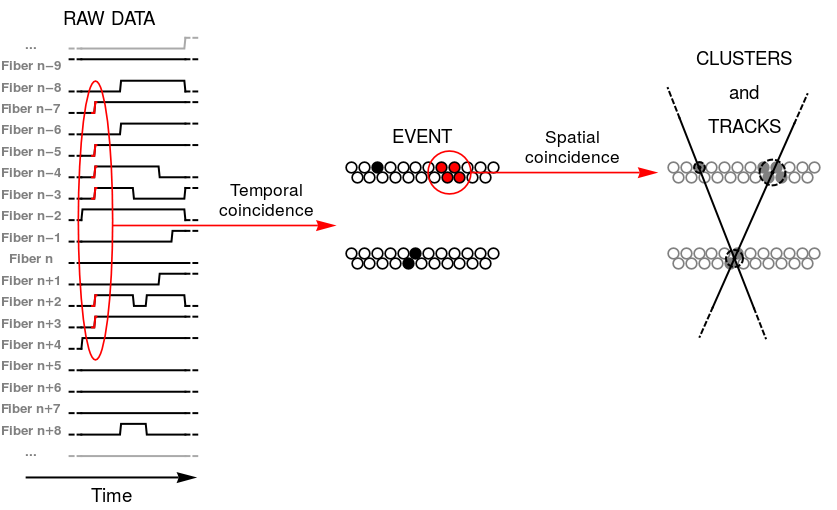}
\caption{General scheme of the tracking algorithm employed in FACT. Rising edges are collected following time coincidence criteria to create events. Adjacent fibers within events are then collected into clusters which are fitted to generate tracks.}
\label{TrackingScheme}
\end{figure}

To discuss the reconstruction of a track in FACT, it is convenient to fix a cylindrical coordinate system having its axis coordinate $z$ coincident with the detector's axis, $r$ as its radial and $\phi$ as its azimuth coordinates.\\
The vertex reconstruction procedure begins with the identification of \emph{events}, defined as the collection of rising edges in temporal coincidence. Since we have chosen \bias{} values which ensure low noise levels (see section \ref{S:Equalization}), noise rejection is not critical. As a result we can employ a coincidence window of $\SI{10}{\nano s}$ (two FPGA clocks), which accounts for tolerances in the trigger signal discrimination and in the synchronization of the FPGA clocks without introducing harmful levels of noise.
Every instance of two successive clocks containing at least two rising edges is considered to be an \emph{event}.
After a list of all the observed \emph{events} has been compiled, it is used to build \emph{clusters}. Within a \emph{cluster}, the activation of spatially adjacent fibers are grouped together and translated into the activation of a ``meta-fiber'' having $r$ and $z$ coordinates given by the barycenter of the $r$ and $z$ coordinates of the fibers that constitute the cluster.

\begin{figure}[htbp!]
    \includegraphics[width=.9\linewidth]{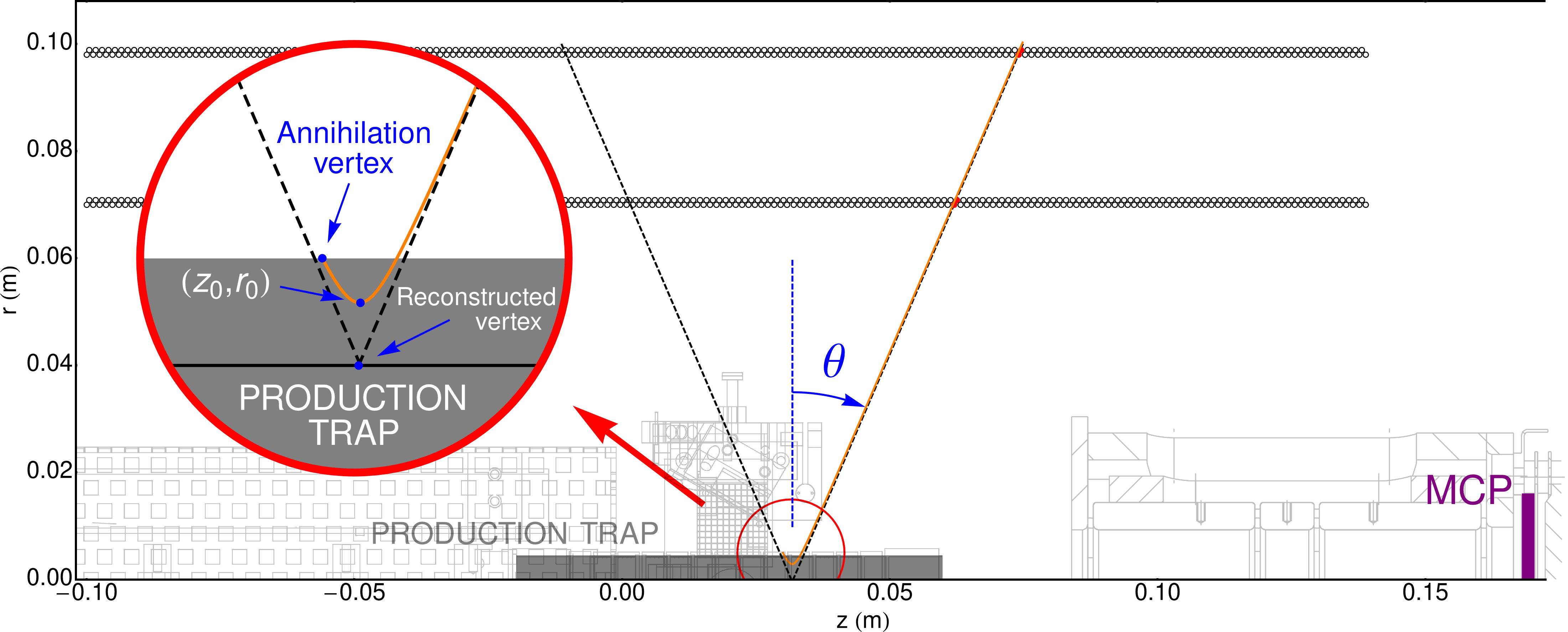}
    \caption{Projection of the straight trajectory originating from an annihilation taking place on the trap wall. The parameters $r_0$, $z_0$ and $\theta$ of the hyperbola given in formula \ref{eq:hyp-r-sqrt} are illustrated along with the position of the annihilation and the reconstructed vertex. The asymptotes of this hyperbola are shown in dashed black.}
    \label{HyperbolaParameters}
\end{figure}

Pairs of clusters situated in different superlayers are used to fit a track in the FACT detector. 
Given the scale of the FACT detector and the energies of the MIPs released after the \antip annihilation, their trajectories will be only marginally affected by the magnetic field of the \unit[1]{T} magnet so that we can assume those trajectories to be straight lines (when ignoring multiple-scattering).  
If we consider a straight trajectory in the three-dimensional space and project it into the $\widehat{rz}$ plane it will result into a hyperbola with the analytical form \cite{HackstockMaster}:

\begin{equation}\label{eq:hyp-r-sqrt}
    r(z) = r_0 \sqrt{1+\frac{(z-z_0)^2}{r_0^2\tan{\theta}^2}},
\end{equation}
where $r_0$, $z_0$ are the coordinates of the minimum of the hyperbola and $\theta=\Delta z/\Delta r$ is the inclination angle of one of the asymptotes. 
This curve cannot be fitted relying solely on two points (one cluster on each superlayer). However we can exploit the fact that the hyperbola's asymptotes will always intersect the $r=0$ axis; therefore, as the vertex of the hyperbola gets closer to the axis, the arms will get closer to being straight lines. Because of the limited size of the production trap the annihilation vertexes will be situated within $\SI{5}{mm}$ of the trap's axis; we therefore opted for approximating the hyperbolae with straight trajectories in the $\hat{rz}$ plane taking their intersection with the $r=0$ axis as the $z$ position of the annihilation vertex (Fig.~\ref{HyperbolaParameters}). Monte Carlo simulation reproducing the geometry of the FACT detector shows that this choice introduces a typical error on the reconstructed $z$ position with respect to the actual $z$ position of the vertex of the underlying hyperbola of \SI{0.9}{\milli m} for vertexes located on the trap axis, \SI{1.3}{\milli m} for vertexes located in the trap volume and 
\SI{2}{\milli m} for vertexes located on the trap walls. Those numbers are within the initial requirements for the detector to achieve a good plasma diagnosis (\unit[2]{mm} corresponding to less than half the width of an electrode), and the monitoring of antihydrogen and beam formation. 

\section{Detector Efficiency}
\label{S:efficiency}

We have developed a method to produce unbiased efficiency estimators for each of the 794 individual FACT fibers\footnote{In the context of the efficiency measurement, ``fiber" describes the entire detection element composed of the scintillating fiber, the clear fiber, the MPPC and the amplification and discrimination electronics.} without having to rely on an external reference and instead using solely the tracking capabilities of the detector itself.
After reconstructing a track we can list all fibers that are crossed by such track: we divide these fibers into two categories based on whether they recorded a rising edge at the track time or not. Fibers which have indeed lit up are named \emph{hit} while the ones which remained silent are labeled \emph{miss}.
If our reconstruction capability was independent of the fibers performance, a fiber's efficiency $\epsilon$ could be determined by counting \emph{hits} and \emph{misses} for a specific fiber, and would be given by

\begin{equation}
    \epsilon ~=~ \frac{N_{\mathrm{hits}}}{N_{\mathrm{hits}} ~+~ N_{\mathrm{misses}}}.
\end{equation}

\begin{figure}[!htbp]
\centering
\begin{subfigure}[t]{.3\textwidth}
  \adjincludegraphics[width=\linewidth]{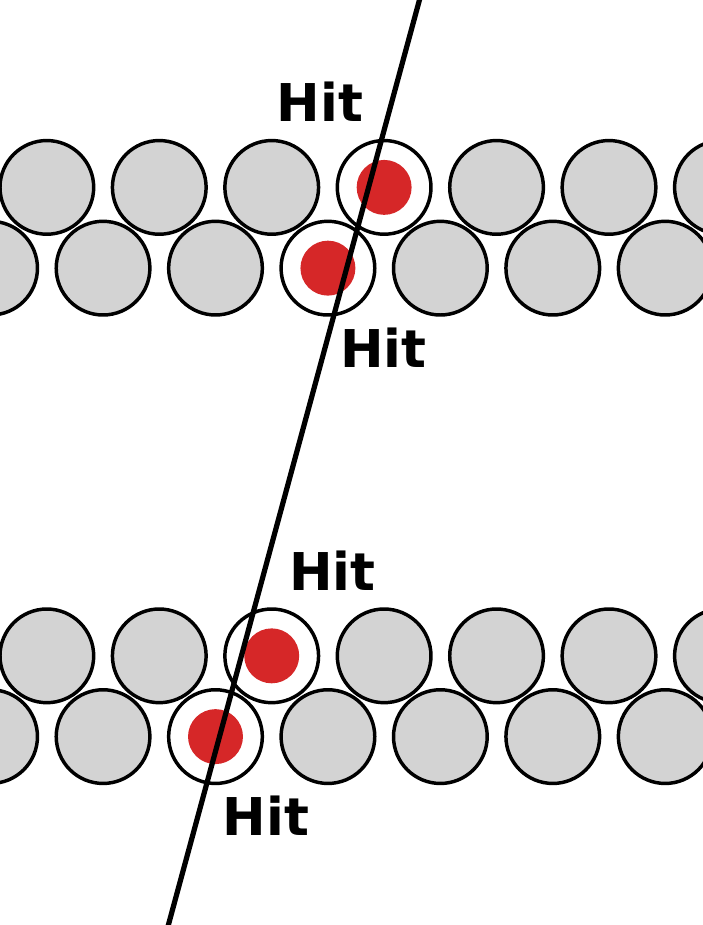}
  \caption{All four fibers \num{4} are firing and are counted as \emph{hits}.}
  \label{fig:4hits}
\end{subfigure}~~
\begin{subfigure}[t]{.3\textwidth}
  \adjincludegraphics[width=\linewidth]{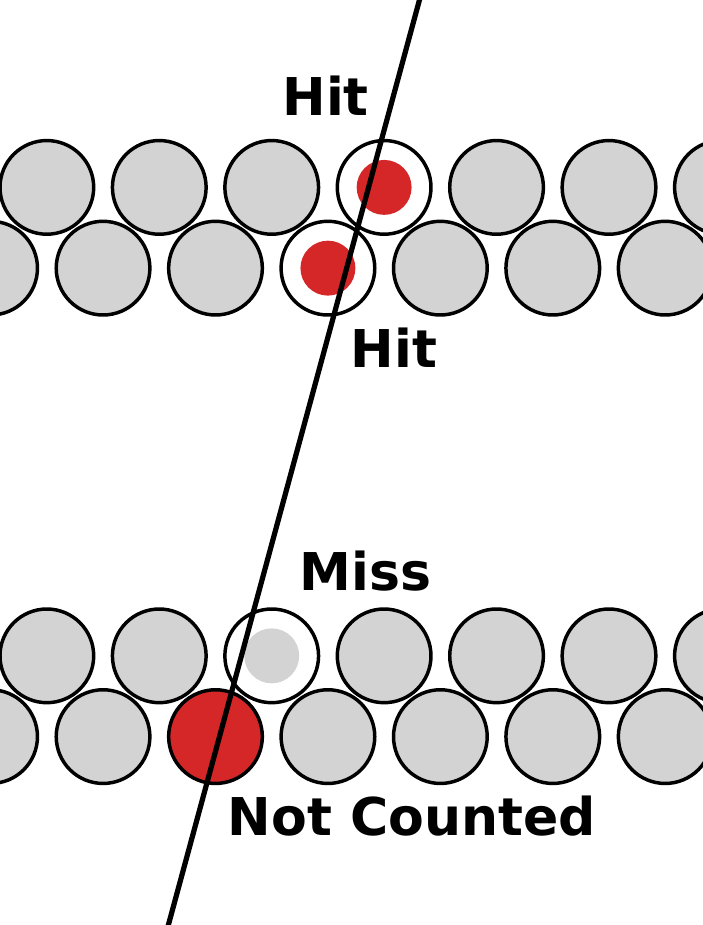}
  \caption{One fiber is missing, the one firing on the same superlayer is not counted as a \emph{hit}.}
  \label{fig:3hits}
\end{subfigure}~~
\begin{subfigure}[t]{.3\textwidth}
  \adjincludegraphics[width=\linewidth]{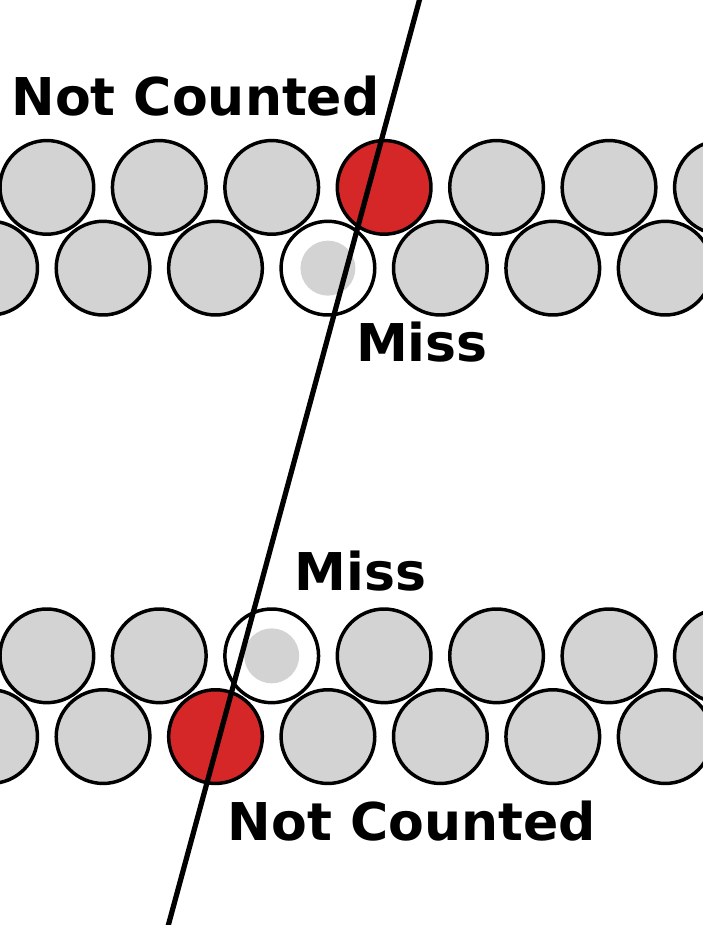}
  \caption{In a two-fiber track, only the missing fibers are counted.}
  \label{fig:2hits}
\end{subfigure}
\caption{Illustrations of the three different hit and miss configurations for tracks  recorded by FACT. The fibers firing are colored in red, those which are counted as ``hit'' are additionally circled in white. The fibers not firing are colored in grey, those which are counted as ``miss'' are additionally circled in white. As it is sufficient for a track reconstruction to have one fiber firing on each superlayer, all three scenarios represent successfully reconstructed tracks. Figures (\subref{fig:3hits}) and (\subref{fig:2hits}) show only one of the possible scenarios for a three and two hits configuration, all the other cases are permutations of the shown ones.}
\label{fig:threecases}
\end{figure}

However, since inefficient fibers can cause the reconstruction of some tracks to fail, the efficiency estimator of a single fiber has to be carefully chosen.
A fiber lighting up can only be considered as a \emph{hit} if the track responsible for this hit could have been reconstructed without considering this fiber. This implies that a cluster consisting of a single fiber on a superlayer cannot be counted towards the \emph{hits}.  
Figure \ref{fig:threecases} illustrates the labelling procedure with three different ``hit-miss" configurations \cite{HackstockMaster}.

We have verified the procedure using a Monte-Carlo simulation which confirmed that the method yields a suitable estimator of the detector efficiency.\\
Two independent measurements based on antiproton annihilations on a downstream detector (a Microchannel plate or MCP, see section \ref{S:Antiprotons}) and on the trap wall led to compatible results giving a mean fiber efficiency of $0.458 \pm 0.019$. From Monte-Carlo simulations we estimate that the fiber cladding and the geometrical arrangement of the fibers are responsible for a loss of about \unit[25 to 30]{\%} of efficiency. 
Given the mean fiber efficiency obtained, and since the observed efficiencies are similar across the detector, its average track reconstruction efficiency can be computed :
\begin{equation}\label{eq:track-eff}
    \overline{\epsilon}_{track}(\overline{\epsilon}) = (1-(1-\overline{\epsilon})^2)^2,
\end{equation}
which yields a value of around \unit[50]{\%}.
This figure, together with the solid angle coverage of the detector ($\unit[2\pi]{str}$), can be used to estimate the signal yield per antihydrogen annihilation. Since on average three pions are produced by every single \antih annihilation, FACT will record on average 0.75 tracks per \antih.

\section{Typical signatures}
\label{S:Signatures}

\subsection{Antiprotons}
\label{S:Antiprotons}

We tested the antiproton detection capability of the FACT detector, along with the reconstruction of axial positions of annihilation vertexes, using \antip manipulation procedures that we denote as \emph{radial release} and \emph{axial (MCP) release}.
An antiproton radial release consists in the deliberate loss of antiproton confinement induced by the manipulation of the segmented electrodes' potentials (rotating wall) with a phase and frequency that induces instabilities in the trapped plasma. A typical radial release liberates $1.5\times10^5$ antiprotons within \SI{100}{\milli s}. An axial release consists in the deliberate loss of longitudinal confinement obtained by rising the potential of the electrodes situated downstream of the production trap, which in turn causes the trapped \antip to annihilate on an MCP located downstream of the production trap. This procedure releases about the same amount of antiprotons in \unit[500]{ms}.

\begin{figure}[htbp!]
    \includegraphics[width=.9\linewidth]{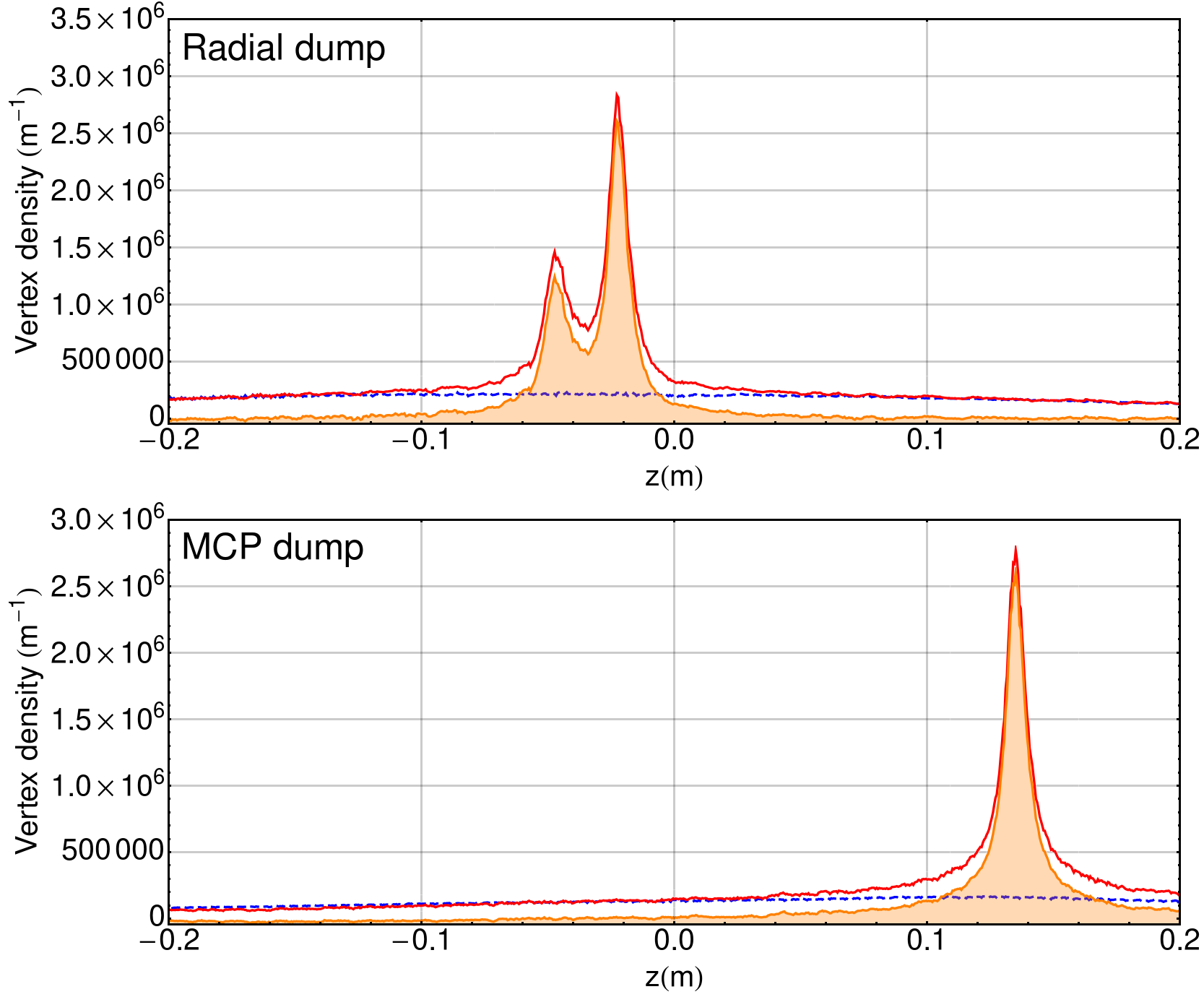}
    \caption{Distribution along the $z$ axis of vertexes reconstructed from a controlled radial release (above) and an axial release (below). The distribution of all reconstructed vertexes, that comprises the \emph{trackable} signal and an \emph{untrackable} background, is shown in red. The dashed blue lines indicate the \emph{untrackable} background as reconstructed via Monte Carlo simulation and in filled-in orange the background-subtracted trackable signal. The peaks in the upper panel are consistent with the position of the upstream and downstream electrodes in between which the \antip~plasma is held during this particular procedure. The peak in the lower panel shows the position of the MCP onto which the antiprotons are released.}
    \label{fig:PBarDump}
\end{figure}

The signal recorded during radial and axial realeases contains rising edges whose coincidence in time indicate a common cause of the signal. Hence coincidences of rising edges can be fitted with tracks, allowing the reconstruction of the axial position at which the annihilation took place. We denoted this kind of signal a \emph{trackable} signal. If we try to reconstruct vertex locations within this signal, we find the resulting distribution to be superimposed onto a background caused by randomly occurring coincidences, which we call \emph{untrackable} background. We can precisely assess the shape of the \emph{untrackable} background by running a Monte Carlo simulation in which we pair random fiber hits from the recorded data and treat them as if they had occurred in coincidence to generate tracks. After reconstructing such expected background we can subtract it from the recorded data, as shown in figure \ref{fig:PBarDump}.

During a radial release, most of the \antip are expected to annihilate in the proximity of the electrodes holding the plasma inside of the trap as observed by the  ATHENA collaboration using similar release procedures \cite{TwoProngs}. The two peaks shown in the top most panel of figure \ref{fig:PBarDump} correspond to the position of the two ends of the stack of electrodes holding the plasma at $z = -\SI{6}{\centi m}$ and $z = 0$.  

For an axial release the expected excess of annihilation vertexes at the position of the MCP is visible in the lower panel of figure \ref{fig:PBarDump}. The FWHM of the peak has been measured to be \SI{7}{\milli m}. The worsened resolution by a factor of $\sim 2$ (compared to the $\sigma <\unit[2]{mm}$ resolution for antiproton annihilations inside the trap) is due to the MCP being located outside of FACT, which causes the reconstructed tracks to have a high value of $\theta$. If tracks with $\theta < 30^\circ$ are selected, the expected resolution at the detector center is recovered. One should note as well that the axial position of the MCP with respect to the trap when held at room temperature (as indicated in figure~\ref{HyperbolaParameters}) is different from the one when the system is cold, leading to the roughly \unit[1.5]{cm} difference in position.

\begin{figure}[htbp!]
    \includegraphics[width=\linewidth]{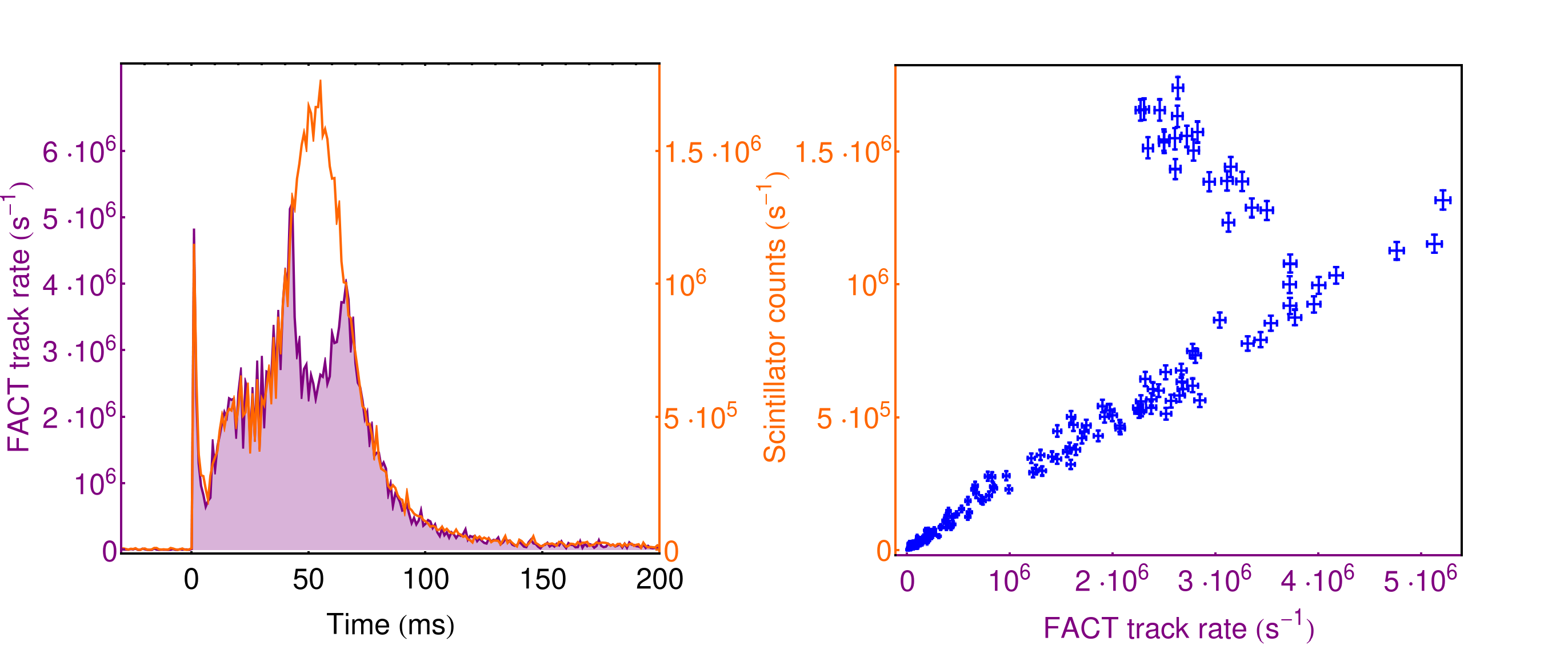}
    \caption{Left: the rate of tracks recorded by FACT during a radial release (in filled-in purple) superimposed with the activity recorded by the \AEgIS scintillating detectors (in orange); at \SI{50}{\milli s} FACT saturates and the rate of recorded tracks drops. Right: the same data is represented as a scatter plot, showing the limit of the FACT detector before saturating to be around $4\times10^6$ tracks/s.}
    \label{fig:Boomerang}
\end{figure}

Antiproton releases can also be used to assess the saturation limitations of the detector. The leftmost panel of figure \ref{fig:Boomerang} shows the rate of tracks recorded by the FACT detector during an intense radial release procedure superimposed to the activity recorded by a set of scintillating detectors~\cite{AegisProposal} installed around the experimental apparatus. Since the scintillating detectors can record a higher signal rate than FACT, they can be used to assess the saturation rate of FACT. The left panel of figure~\ref{fig:Boomerang} shows that at around  \SI{50}{\milli s}, for this particular antiproton release, the track rate in FACT drops. The scatter plot (scintillator count rates versus FACT track rate) in the right panel helps visualizing that the loss-free operation of FACT is achieved up to a maximum rate of $\unit[4\times10^6]{tracks/s}$. The points beyond this rate correspond to spikes in the activities which can appear at the beginning of the saturation period.

\subsection{Positrons}

\label{S:Positrons}

\noindent 
Positron bunches are produced using the \AEgIS positron system \cite{AEGpos}. Briefly, positrons emitted by a $^{22}$Na source are slowed down by a solid Ne moderator \cite{Mills86} and then prepared by a Surko-style trap \cite{Danielson15} and accumulator. The number of positrons in each bunch is proportional to the number of pulses from the Surko trap stored in the accumulator. When such a bunch is shot from the accumulator onto the conversion target, the entire FACT detector saturates. Following the initial burst the detector is blinded, with all of the fibers staying over threshold, for a period ranging from 0.4 to \SI{1}{\micro s}.

\begin{figure}[htbp!]
    \centering
    \includegraphics[width=.7\linewidth]{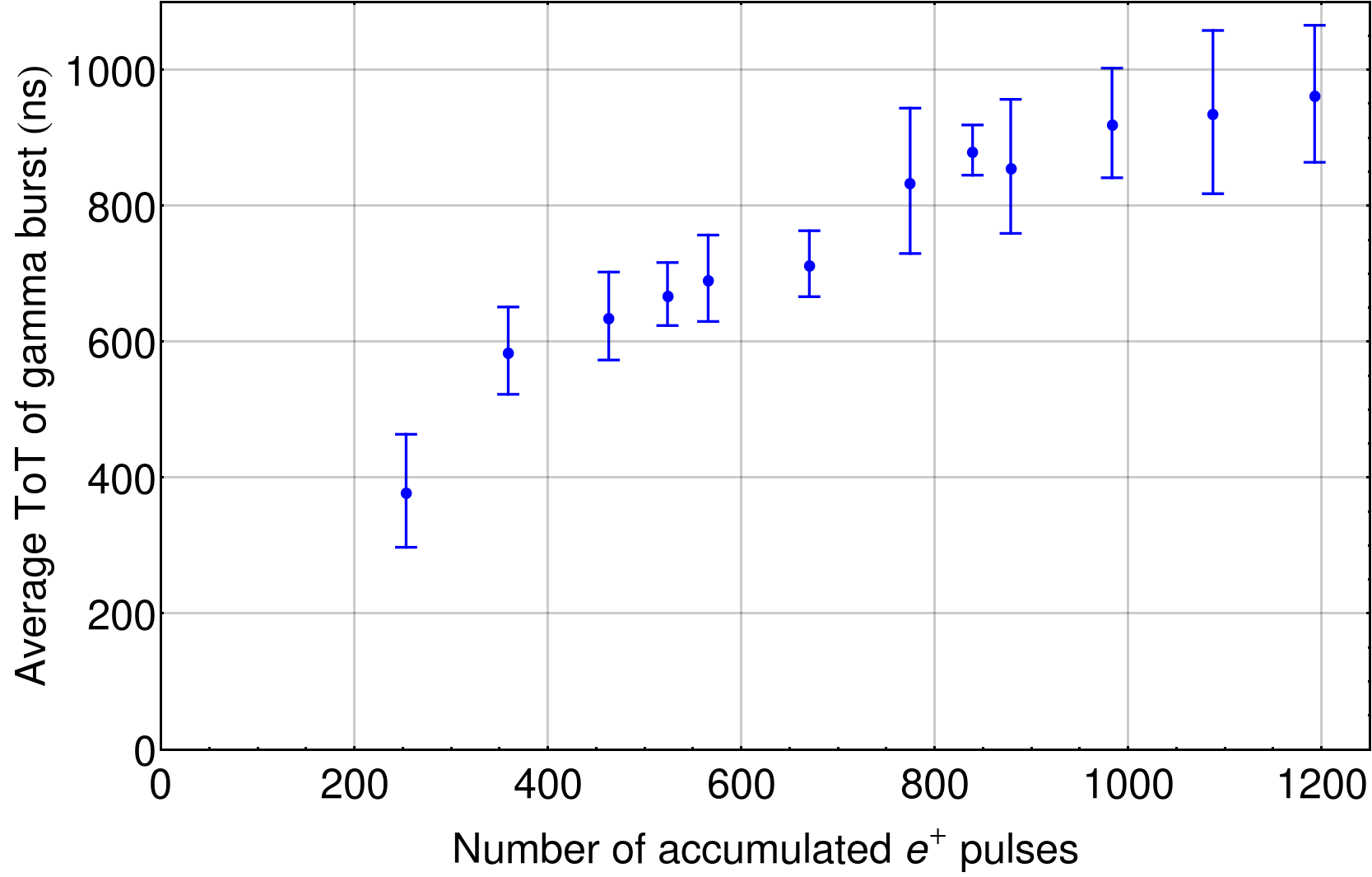}
    \caption{Average time-over-threshold of the first event following the positron burst as a function of the number of accumulation cycles constituting the positron shot. The number of cycles, as long as the saturation of the positron accumulation trap is not reached, is a good measurement of the number of positrons present in a single shot. As it can be seen, from 400 accumulated shots onward, the duration of the detector blindness is roughly linear in the number of positrons shot.}
    \label{AlexanderPlot}
\end{figure}

\begin{figure}[htbp!]
    \centering
    \includegraphics[width=.7\linewidth]{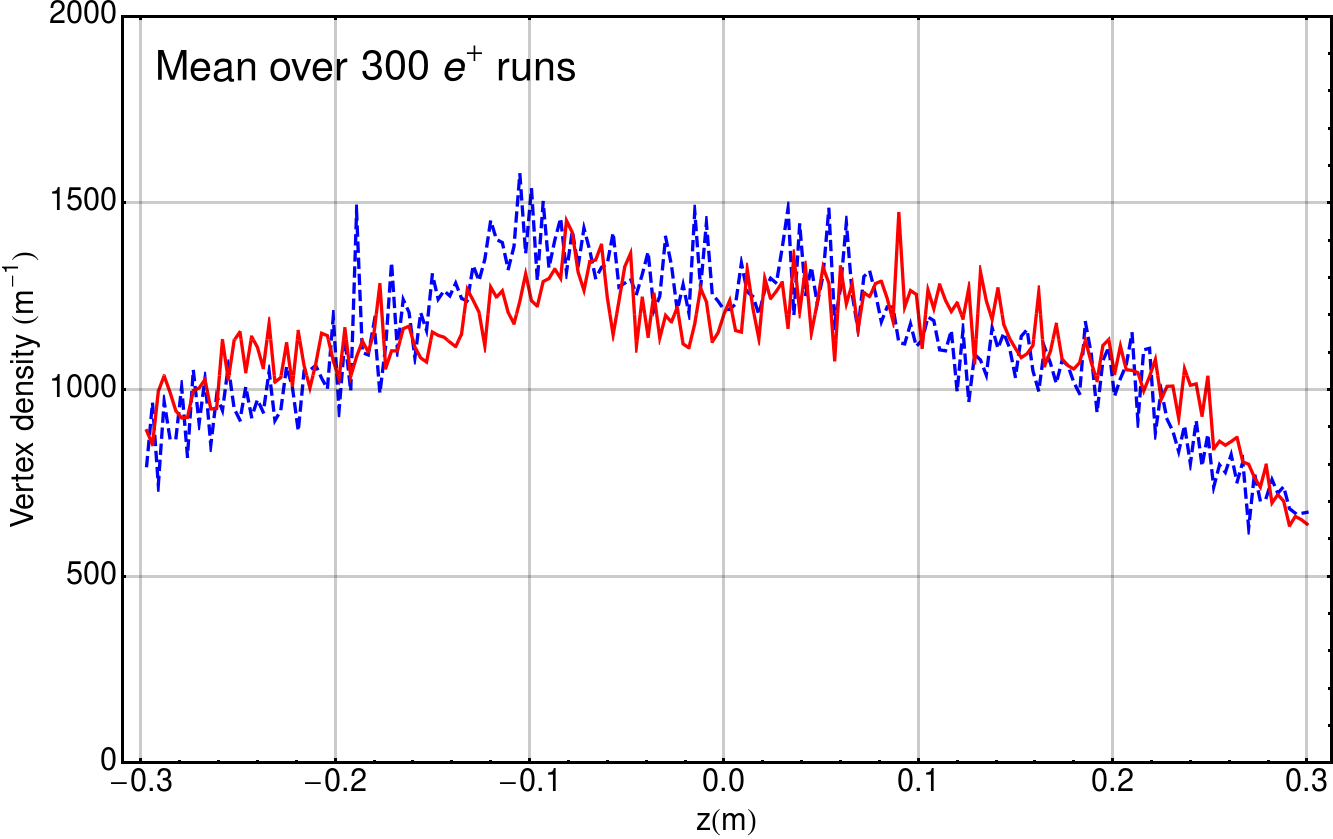}
    \caption{Position along the $z$ axis of the reconstructed vertexes recorded in FACT  within $\unit[5]{\mu s}$ after the positron shot (in red). As the observation window is much smaller than that available for \antip release procedures, 300 runs had to be combined together to obtain a distribution. In dashed blue the expected \emph{untrackable} component computed from Monte-Carlo (see \S\ref{S:Antiprotons}) is shown and can describe the entirety of the signal.}
    \label{fig:EPlusTracking}
\end{figure}

The duration of the blindness region at the fiber level is dependent on the amount of gamma rays hitting the fiber. This causes an uneven recovery: we have observed instances in which the fibers located at the edge of the detector (covering a smaller solid angle with respect to the positronium converter) recovered after \SI{400}{ns} while the fibers located in the center of the detector required around \SI{1.2}{\micro s} to recover. The dependency of the blindness period is roughly linear in the number of injected positrons (Fig.~\ref{AlexanderPlot}).
The blindness period is followed by a tail caused by afterpulses in the MPPC detectors. This signal, see figure \ref{fig:EPlusTracking}, is by its nature \emph{untrackable} as  is confirmed by a Monte Carlo simulation performed in a similar way as exposed in section \ref{S:Antiprotons}.

\subsection{Typical \antih production cycle signature.}

An antihydrogen production cycle begins with the capture of several \antip shots from the AD beamline. Antiprotons are then cooled, compressed and transferred into the production trap. A bunch of positrons is then sent onto a silicon converter and the excitation laser is pulsed initiating the chain of events culminating in the production of antihydrogen. We set as $t=0$ the time at which the positron bunch impacts onto the converter target. At $t = -\SI{3}{\milli s}$ FACT receives a trigger and starts recording the activity inside of the production trap.
\begin{figure}[htp]
\centering
\includegraphics[width=\textwidth]{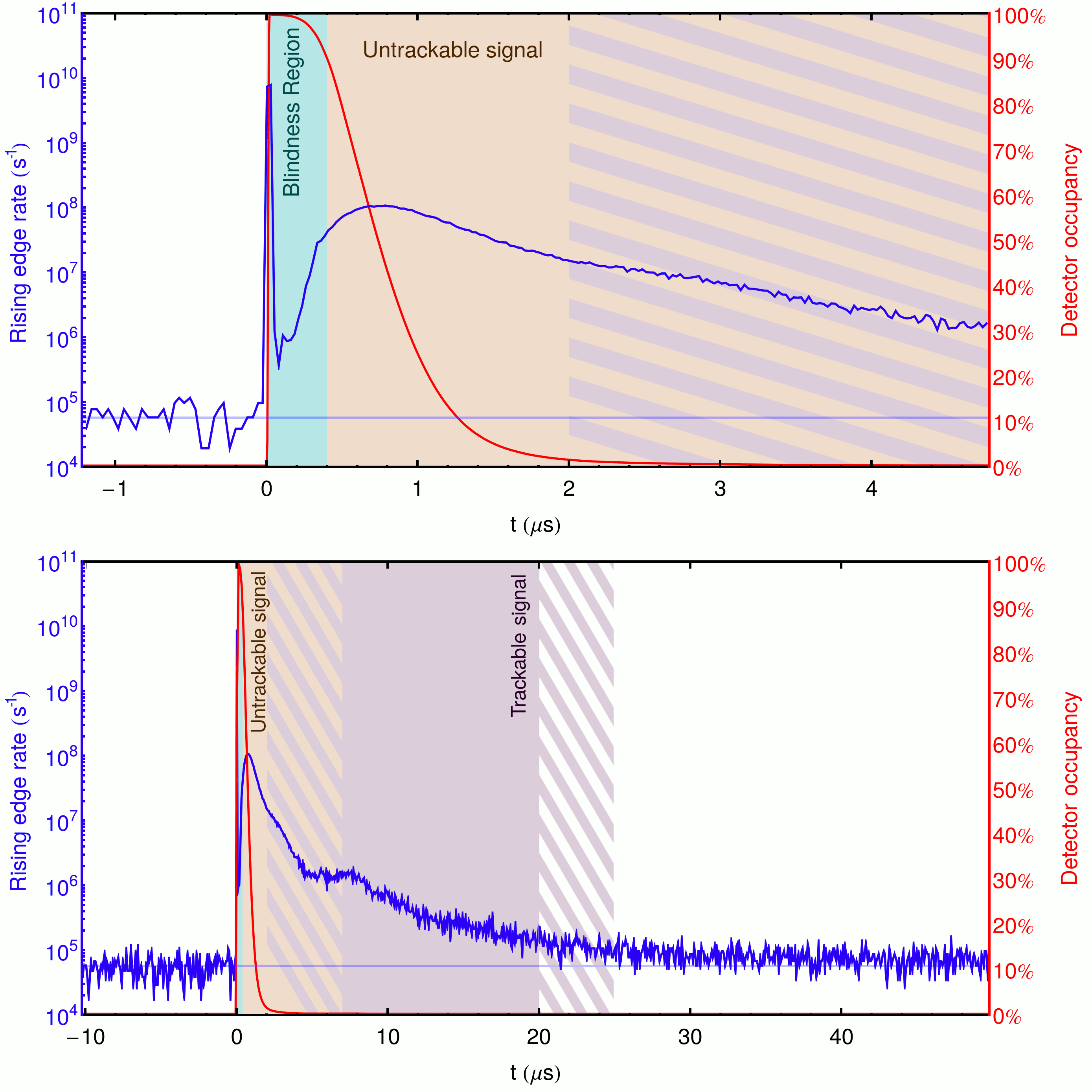}
\caption{Typical shape of the activity in FACT during an \antih production cycle. The two frames show the same curve at different magnification levels. To obtain these profiles, 1027 different \antih production cycles have been averaged together. The blue line shows the global rate of rising edges in the detector, the red curve which fraction of the detector's fibers are above threshold at any single moment. The horizontal blue line indicates the dark count rate of $\unit[50]{s^{-1}}$ per fiber. The first $\unit[0.4]{\mu s}$ after the positron dump is a period of complete detector blindness. This period is followed (up to $\sim\unit[2]{\mu s}$) by a partial recovery of the detector where fibers furthest away from the positron annihilation point exhibit periods of no activity while the central fibers are still on. This period consists fully of untrackable signal as shown in the upper panel of Fig.~\ref{Trackability}. Following this region, is a high activity region including a mix of untrackable and trackable signals (bi-colored hatched region). From  $\sim\unit[7]{\mu s}$ the signal is dominated by trackable signals as can be seen from the lower panel of Fig.~\ref{Trackability}. After  $\sim\unit[20]{\mu s}$ the signal becomes again a mix of untrackable and trackable signals (last hatched region in the lower panel) until the signal is consistent again with the average dark count rate of the detector.}
\label{TypicalShot}
\end{figure}
Since the \antip losses are much more feeble than the dark rate of FACT, in the $\unit[-3]{ms} < t < \unit[0]{s}$ range the detector is substantially silent with the dark counts ($\sim\unit[50]{s^{-1}}$/fiber) dominating the detected signal as indicated by the horizontal line in figure \ref{TypicalShot}. At $t = 0$ the detector is completely blinded by the $\gamma$ rays resulting from positron annihilations in the converter target and, shortly after, by the annihilation of the produced positronium.
 Antihydrogen formation procedures were performed with a reduced number (500) of accumulated pulsed per positron shot, corresponding to \positroncount positrons, in order to mitigate the effect of the MPPCs saturation at early times. In this configuration, the positron shot is followed by a period of complete blindness of the entire detector lasting typically  $\unit[0.4]{\mu s}$ 
during which all of FACT fibers stay above their threshold. This is followed by a transition period (up to $\sim \unit[2]{\mu s}$) where the fibers furthest away from the converter target recover from the blindness period but exhibit a high activity while the central fibers are still blinded. Vertex reconstruction (see upper panel of figure \ref{Trackability}) shows this period to be fully dominated by \emph{untrackable} signals, similar to the ones which were obtained analyzing the period shortly after the $e^+$ shot.
From $\sim \unit[2]{\mu s}$ to roughly $\sim \unit[5]{\mu s}$, all fibers have recovered from the blindness period but the global rate of rising edges remains between $10^6$ and $\unit[10^8]{s^{-1}}$, dominated by \emph{untrackable} signals.
 This large signal precludes the efficient detection of a \emph{trackable} signal which would appear in the first $\sim\unit[5]{\mu s}$ after the positron shot. If those were signals from antihydrohen atoms formed in the center of the trap and annihilating on the inner trap wall, the maximum \antih temperature which could be detectable, taking into account the geometry of the trap, would be $\sim\unit[50]{K}$. After this activity has faded out, in the region leading up to roughly $\SI{20}{\micro s}$, an excess of activity with respect to the dark rate still manifests. The rate of reconstructed vertexes is typically between $10^5$ and $\unit[10^6]{s^{-1}}$. The $z$ distribution of the vertexes, shown in the lower panel of figure \ref{Trackability}, indicates that the signal consists mostly of a \emph{trackable} component, similarly to that which was recorded during \antip release procedures, and shows a peak at the $z$ position at which the \antip plasma is held.

\begin{figure}[htp]
\centering
\includegraphics[width=0.7\textwidth]{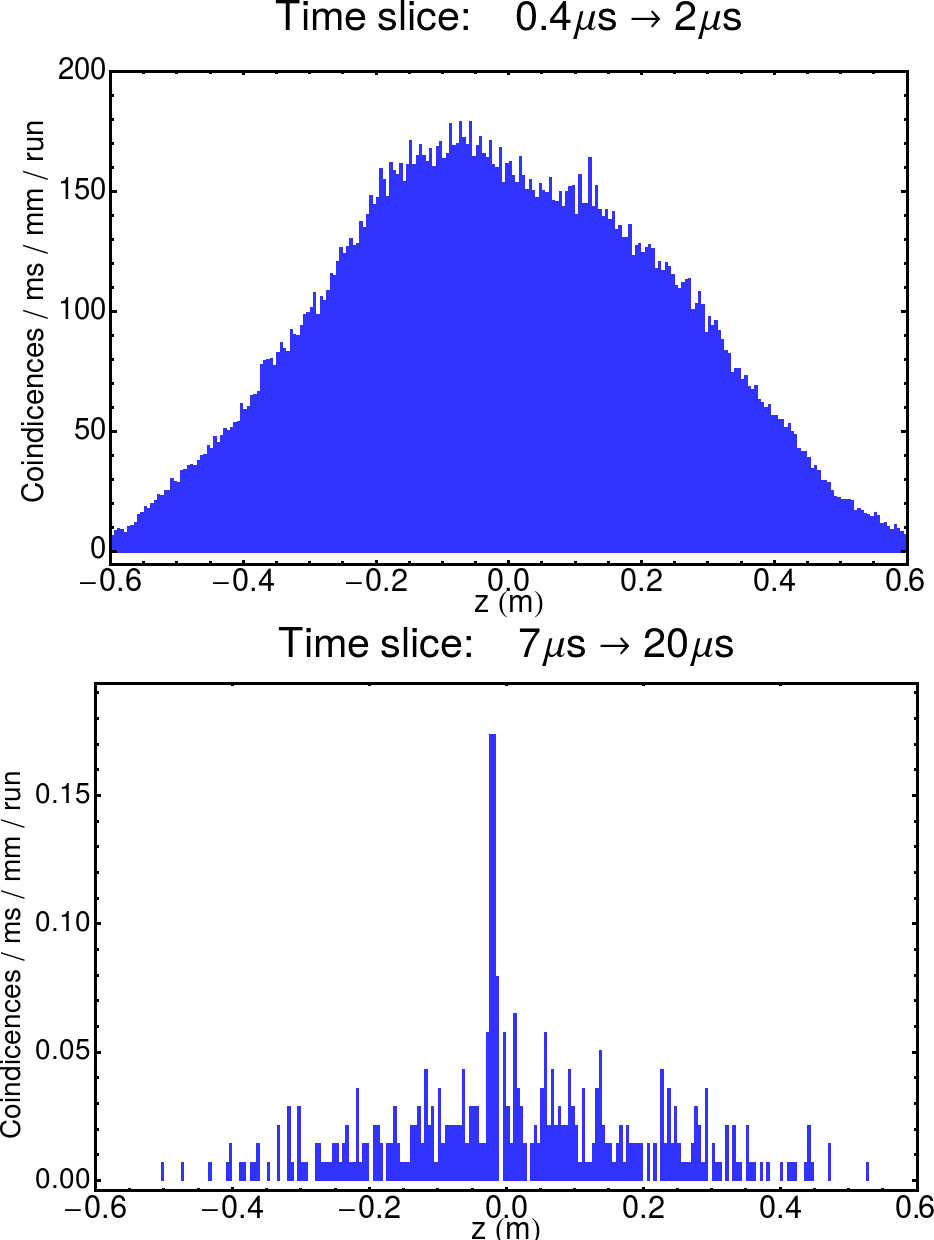}
\caption{
Distribution of vertexes for the signal shown in figure \ref{TypicalShot} for two different time slices highlighting the typical z-distribution for untrackable (top) and trackable signals (bottom).}
\label{Trackability}
\end{figure}

\section{Conclusions and outlook }
\label{s:conc}

In this work we have described the main characteristics of the FACT detector as well as a series of developments which led to the successful operation of FACT during \antih production runs in the \AEgIS experiment. Thanks to a read-out system relying on a fast Ethernet link and a dedicated control software, the automation of the data acquisition and a synchronous transfer of the recorded data was implemented which in turn allowed close to arbitrarily-long acquisitions.  An equalization procedure of the 794 fibers was developed which reduced the detector equalization time to a few seconds, a short enough duration to allow re-equalization between each \antih run.
The dependency of the MPPC performance on temperature has been characterized,  
and mitigated thanks to a thermal drift compensation system, which together with the fast re-equalization procedure drastically increased data quality. A tracking algorithm was developed and employed to bootstrap the determination of the fiber efficiency, which was found to be roughly \unit[45]{\%}. The vertex reconstruction capability of the detector was tested employing \antip release procedures. The lack of radial sensitivity leads to an axial resolution between $\sim\unit[1]{}$ and $\unit[2]{mm}$ depending on the radial position of the annihilation, which is sufficient to monitor the annihilation of antihydrogen atoms travelling toward the envisioned deflectometer apparatus. The blinding of the detector caused by the injection of positrons into the apparatus was characterized. The limitations in the first $\unit[5]{\mu s}$ after the positron pulse, precluding the detection of \antih with a temperature above $\sim$\unit[50]{K}, is linked to the response of the MPPCs. We have considered the possibility of reducing the blindness region by injecting a pulsed signal in the MPPC \bias{} voltage, timed so that the bias is null when the positron shot reaches the converter target. Preliminary tests using this technique have already shown promising results and are envisioned to increase the capabilities of the detector.\\ 
These investigations showed that FACT is a powerful plasma diagnostic tool and is capable of detecting cold \antih annihilating on the inner trap surface, owning its capabilities in part to its low noise, good timing and millimetric axial position resolution.

\section*{Acknowledgment}
This work was financyally supported by Istituto Nazionale di Fisica Nucleare;
the CERN Fellowship programme and the CERN Doctoral
student programme; the Swiss National Science
Foundation Ambizione Grant (No. 154833);
a Deutsche Forschungsgemeinschaft research grant; an excellence initiative of Heidelberg University;
the European Union's Horizon 2020 research and innovation programme under the Marie Sklodowska-Curie grant agreement No. 754496 - FELLINI;
Marie Sk\l{}odowska-Curie Innovative Training Network Fellowship of the European Commission's Horizon 2020 Programme (No. 721559 AVA);
Marie Sk\l{}odowska-Curie Cofund Action research and innovation Programme of the  European Union’s Horizon 2020 (No. 754496);
European Research Council under the European Unions Seventh Framework Program FP7/2007-2013 (Grants Nos. 291242 and 277762);
European Union's Horizon 2020 research and innovation programme under the Marie Sklodowska-Curie grant agreement ANGRAM No. 748826;
Austrian Ministry for Science, Research, and Economy;
Research Council of Norway; Bergen Research
Foundation; John Templeton Foundation; Ministry
of Education and Science of the Russian Federation
and Russian Academy of Sciences and the European Social Fund within the framework of realizing the project,
in support of intersectoral mobility and the European
Social Fund within the framework of realizing Research
infrastructure for experiments at CERN, LM2015058.
We thank the mechanics and electronics workshops of the Physik-Institut (Zurich University) and of the Laboratory of High Energy Physics (Bern University).

\section*{Bibliography}

\bibliographystyle{model1-num-names}

\bibliography{sample}

\end{document}